\documentclass[11pt,a4paper]{article}
\usepackage{jcappub}
\usepackage{float}
\usepackage[T1]{fontenc} 
\usepackage[english]{babel}
\usepackage{graphicx}
\usepackage{siunitx}
\usepackage{amsmath,latexsym}
\usepackage{amssymb}
\usepackage{multirow}
\usepackage{enumerate}
\DeclareSIUnit \belm {Bm}
\DeclareSIUnit{\Sample}{S}
\DeclareSIUnit\year{yr}
\usepackage[textwidth=2cm]{todonotes}
\graphicspath{{.},{./figures/}}

\newcommand{\dd}{\mathrm{d}}

\usepackage{pgf}
\usepackage{soul}
\usepackage{xcolor}
\definecolor{lightgreen}{HTML}{B7F774}
\definecolor{dblue}{HTML}{0000BB}
\definecolor{lhred}{HTML}{FF0000}

\sethlcolor{yellow}

\title{First results from the WISPDMX radio frequency cavity searches for hidden photon dark matter}

\author[a]{Le Hoang Nguyen}
\author[b,a]{Andrei Lobanov}
\author[a]{and Dieter Horns}
\emailAdd{le.hoang.nguyen@uni-hamburg.de}
\emailAdd{andrei.lobanov@mpifr-bonn.mpg.de}
\emailAdd{dieter.horns@uni-hamburg.de}

\affiliation[a]{Institut f{\"u}r Experimentalphysik, Universit{\"a}t Hamburg, Hamburg, Germany.}
\affiliation[b]{Max-Planck-Institut f{\"u}r Radioastronomie, Auf dem H\"ugel 69, 53121 Bonn, Germany.}

\abstract{The origin of non-baryonic dark matter remains elusive
  despite ongoing sensitive searches for heavy, thermally produced
  dark matter particles.  Recently, it has been shown that
  non-thermally produced vector bosons (sometimes called hidden
  photons) related to a broken U(1) gauge symmetry are among the
  possible WISP (weakly interacting slim particles) dark matter
  candidates.  The WISP Dark Matter eXperiment (WISPDMX) is
  the first direct hidden photon dark matter search experiment probing
  the particle masses within the \SIrange{0.8}{2.07}{\micro\eV} range
  with four resonant modes of a tunable radio frequency cavity and
  down to \SI{0.4}{\nano\eV} outside of resonance.  In this paper, we
  present the results from the first science run of WISPDMX comprising
  \num{22000} broadband spectra with a \SI{500}{\mega\Hz} bandwidth
  and a \SI{50}{\Hz} spectral resolution, obtained during 10-second
  integrations made at each individual tuning step of the
  measurements. No plausible dark matter candidate signal is found,
  both in the individual spectra reaching minimum detectable power of
  \SI{8e-19}{\watt} and in the averaged spectrum of all the
  measurements with the minimum detectable power of \SI{5e-22}{\watt})
  attained for a total of \SI{61}{\hour} of data taking.
  Using these spectra, we derive upper limits on
  the coupling constant of the hidden photon at the levels of \num{e-13} for
  the resonant frequency ranges and \num{e-12} for broadband mass range
  \SIrange{0.2}{2.07}{\micro\eV}, and steadily increasing at masses below
  \SI{0.2}{\micro\eV}.  }

\begin{document}
\maketitle
\keywords{
 Dark Matter \sep  Hidden Photons \sep Resonant Cavity 
}
\section{Introduction}

The standard model of particle physics (SM) has so far withstood all
experimental tests, yet it remains still incomplete.  Apart from the
well-recognized shortcomings of the model such as the apparently
arbitrary choice of parameters and issues related to naturalness
\cite{dine_naturalness_2015}, the SM fails to provide a viable
explanation for dark matter and spectacularly over-predicts dark
energy, which are the two dominating components of the $\Lambda$CDM
(cosmological constant $\Lambda$, cold dark matter) standard model of
cosmology (for some non-standard insights to the $\Lambda$CDM model,
see \cite{scott_standard_2018}).  In many extensions of the standard
model, a hidden (or dark) sector is introduced to provide an
explanation for the missing components observed in our universe
(without violating other bounds). The hidden sector is minimally
coupled via gravity to the common forms of matter; additional
mediators include the so-called Hidden Photons (HPs), an Abelian boson
uncharged under the Standard Model (SM) fields of the visible sector
\cite{okun1982,goodsell_naturally_2009}.  Hidden photons can be
produced during inflation, avoiding thermalization and providing a
relic abundance which is consistent with the estimated dark matter
density in the Universe
\cite{nelson_dark_2011,arias_wispy_2012,graham2016}.  The hidden
vector boson mixes kinetically with the SM photon providing a weak
coupling \cite{holdom1986,abel_kinetic_2008}. The corresponding
Lagrangian (natural units with $\hbar=c=1$ are used) is given by

\begin{equation}
  \mathcal{L} = -\frac{1}{4} F_{\mu \nu}F^{\mu \nu} - \frac{1}{4} X_{\mu \nu}X^{\mu \nu} + \frac{m^2_{\gamma'}}{2} X_\mu X^\mu + \frac{\chi}{2} F_{\mu \nu}X^{\mu \nu} + J_\mu A^\mu
\end{equation}
where $X_{\mu\nu}=\partial_\mu X_\nu - \partial_\nu X_\mu$ and
$F_{\mu\nu}=\partial_\mu A_\nu - \partial_\nu A_\mu$ are the HP's and
SM electromagnetic field tensors respectively, while $J^\mu$ is the
current of electric charges.  The Lagrangian above includes the
coupling between the HP field and photon field via kinetic mixing. The model has two parameters, the mass, $m_{\gamma'}$,
of the boson and the coupling strength, $\chi$, of its kinetic mixing
with the SM photon.  The actual values of $m_{\gamma'}$ and $\chi$ are
related to the structure of the dark sector interactions and
the choice of the mass-generating mechanism (St\"uckelberg or Higgs).
Theoretical considerations provide only loose constraints on possible
choices of the parameters (see e.g., \cite{abel_kinetic_2008,
  goodsell_naturally_2009}), with some preference for the
$10^{-12}$--$10^{-2}$ range of the kinetic mixing coupling
\cite{essig_dark_2013}.

Assuming that the HP is also the main constituent of dark matter,
additional constraints on $m_{\gamma'}$ and $\chi$ can be derived
\cite{nelson_dark_2011,arias_wispy_2012}. When considering a
St\"uckelberg mass generation mechanism, the HP condensate remains
cold and stable against decay and evaporation after its formation
during inflation when choosing values $\chi\lesssim 10^{-11}$ for a
broad range of masses (one should bear in mind that the relic density
produced with the misalignment mechanism depends on the initial value
of the field). 
Further observational and experimental
constraints on $m_{\gamma'}$ and $\chi$ have been summarized in
\cite{arias_wispy_2012}.

A number of these constraints result from experiments employing
resonant cavities for enhancing the detection sensitivity for the
photon signal produced inside the cavity by the incident dark matter
particles (see \cite{bradley2003} and references therein). In this
case, the fractional bandwidth of an individual measurement is
inversely proportional to the signal enhancement which typically
exceeds a factor of $10^4$. For such experiments to cover a sizable
range of particle mass, the cavity must be tuned, and a larger number
of independent, narrow band measurements must be made. The WISP Dark
Matter eXperiment (WISPDMX) described here expands this conceptual
approach by using a tunable large-volume cavity and a broadband
recording apparatus with a total bandwidth of \SI{500}{\mega\Hz}
(\SI{2.07}{\micro\eV}) and a frequency resolution of \SI{50}{\Hz}
(\SI{0.21}{\pico\eV}). This setup makes it possible to combine a set of
tunable resonant measurements made simultaneously at four different
cavity modes with out-of-resonance measurements made essentially over
the entire \SI{500}{\mega\Hz} frequency range.

The relevant physical foundations of WISPDMX measurements are outlined in
Section~\ref{sec1}, focusing on specific aspects related to broadband signal
recording covering multiple resonant modes of the
cavity. Section~\ref{sec:onwispdmx} presents the experimental setup, including
the tuning, frequency calibration, and data acquisition systems and data taking
procedures. The first science run of WISPDMX is summarized in
Section~\ref{sec:analysis}. 
Results of the signal searches and the
corresponding exclusion limits for hidden photon dark matter are presented in
Section~\ref{sec:results} and discussed in a broader context in
Section~\ref{sec:discussion}.

\section{Hidden Photon Search with Resonant Cavity}
\label{sec1}

An electromagnetic resonator can be used for axion dark matter
detection in a \textit{haloscope}-type experiment
\cite{sikivie_experimental_1983} using a hollow resonator or an LC
circuit \cite{arias_extracting_2014}) placed in a strong magnetic
field. Without the magnetic field, the resonator can also be utilized
to search for hidden photons as outlined in
\cite{arias_wispy_2012}. Under a particular assumption that the HP
energy density equals the local dark matter density,

\begin{equation}
\rho_{DM}=\frac{m_{\gamma'}}{2} |\mathbf{X}|^2\,, 
\end{equation}
this leads to an oscillating quasi-stationary electric
field\footnote{The coherence length is given by
  $|\mathbf{p}|^{-1}=\mathcal{O}(\mathrm{km})$ assuming a velocity in
  the halo of \SI{200}{\km\per\second} and a mass
  $m_{\gamma'}=$\si{\micro\eV}} suppressed by a factor $\chi$ which
can in principle be measured in various ways. In the zero-momentum
limit, the hidden photon mass relates to the frequency, $\nu$, of
the oscillations, so that $m_{\gamma'} =
2\pi\nu = 4.13567\,\mu\mathrm{eV}(\nu/1\,\mathrm{GHz})$.

The resulting hidden photon signal, $P_\mathrm{hp}$, in a hollow resonator with a volume, $V$, and
an unloaded quality factor, $Q_0$, determined by the surface losses is given by
\begin{equation}
\label{eqn:power}
  P_\mathrm{hp} = \chi^2 m_{\gamma'} \rho_\mathrm{DM}\, V\, Q_0\, \mathcal{G},
\end{equation}  
where $\mathcal{G}$ represents the geometrical form factor of the
cavity, which expresses an effective volume of the cavity
available for a given resonant mode. For hidden photons,
\begin{equation} 
\label{eqn:formfactor}
   \mathcal{G} = \frac{|\int \dd\,V \textbf{E}^{*cav}(\textbf{x}) \cdot \hat{\textbf{n}} |^2}
{V~\int \dd\,V|\textbf{E}^{cav}(\textbf{x})|^2}\,,
\end{equation}
where $\textbf{E}(\textbf{x})$ is the electric field at the location
$\textbf{x}$ inside the cavity and $\hat{\textbf{n}}$ is the direction
of the HP flux.  Recent studies of the structure formation with a
light vector boson dark matter particle \cite{2012PhRvD..86b1301C}
favor the isotropic distribution of $\hat{\textbf{n}}$, with the
respective average over all directions, $\theta$, given by $\langle
\cos^2(\theta)\rangle=1/3$.  In the simplest case of a pill-box type
cavity, the fundamental transverse magnetic (TM$_{010}$) mode provides
the largest form factor, while it drops substantially for higher order
modes.

The output power, $P_\mathrm{out}$, measured by an antenna inserted
into the cavity is determined by the loaded quality factor, $Q$,
reflecting the antenna coupling, $\kappa$, to the field inside the
cavity, so that
\begin{equation}
 P_\mathrm{out} =  \kappa\, \chi^2 m_{\gamma'} \rho_\mathrm{DM}\, V\, Q\, 
\mathcal{G}\,,
\label{eqn:Pout}
\end{equation} 
up to the critical coupling of the antenna.

\subsection{Broadband Gain}

In haloscope-type searches for light dark matter, at each measurement
step only a narrow frequency range around the relevant resonant
frequency, $\nu_0$, of the cavity is considered. Accordingly, the
signal recording is made over a narrow bandwidth,
$\Delta\nu_\mathrm{rec}\ll \nu_0$, \cite{bradley2003}, and the
sensitivity of each of the measurements is adequately described by the
gain, $g = Q\,\mathcal{G}$, calculated for a given resonant mode at
its respective $\nu_0$. In contrast to that approach, WISPDMX records
an instantaneous bandwidth of \SI{500}{\mega\Hz} which extends over
several resonant modes that are sensitive to the hidden photon signal. This
makes it possible to perform simultaneous measurements with each of
these modes and also to search for an off-resonance signal over the
entire recording bandwidth.

To apply this approach, a broadband gain, $g(\nu)$, of the cavity
needs to be calculated for the entire measured bandwidth, comprising
the cumulative effect of all relevant resonant modes and the
off-resonance response of the apparatus. In this calculation, we
assume that the frequency dependence of the quality factor,
$Q_i(\nu)$, of a particular resonant mode is described by a Lorentzian
profile with the peak at the value at the respective resonant
frequency, $\nu_i$. We also note that, while the quality factor drops
quickly off-resonance, the field configuration and therefore the form
factor $\mathcal{G}_i$ of a given mode remain unchanged. With this,
the broadband gain, $g_i(\nu)$, of a single resonant mode is given by
$Q_i(\nu) \mathcal{G}_i$, and the cumulative, multiple mode response of a cavity
can be written as a sum over all relevant resonant modes:
\begin{equation}
g(\nu) = \sum_i Q_i(\nu) \mathcal{G}_i\,,
\label{eqn:bbgain}
\end{equation}
evaluated over the recorded range of frequencies. Substituting this
term into Eqn.~\ref{eqn:Pout} and recalling that the antenna coupling,
$\kappa$, also varies with frequency, gives
\begin{equation}
P_\mathrm{out}(\nu) = \kappa(\nu)\, \chi^2 m_{\gamma'}(\nu)\, \rho_\mathrm{DM}\, V\, g(\nu)\,
\label{eqn:sum_gain}
\end{equation} 
for the output power measured at a frequency $\nu$. Thus, by measuring the power
$P_\mathrm{out}(\nu)$ over a broad range of frequency, it is possible to
constrain the energy density of HP over the respective range of particle mass,
$m_{\gamma'}$.

\subsection{WISP Dark Matter Signal in Broadband Searches}
\label{section:signaWidth}

The total energy of a non-relativistic hidden photon with a mass,
$m_{\gamma'}$, is given by the sum of the total rest mass energy and
kinetic energy of the particle:

\begin{equation}
E = m_{\gamma'} c^2 + \frac{1}{2} m_{\gamma'} \bar{v}_{\gamma'}^2\,.
\end{equation}
Interpretation of measurements made with a haloscope depends therefore
on the knowledge of the local density and velocity distribution of
hidden photons constituting the dark matter In the standard halo model
(SHM), the Milky Way halo is described as a self-gravitating,
isothermal, and pressureless sphere of particles (see \cite{evans2019}
and references therein). We will not consider here various higher
order structures such as streams and caustics
\cite{duffy2008,vogelsberger2011} which may be present in the halo and
potentially even lead to enhancing the dark matter signal in a
haloscope \cite{lentz2017}.  The local density of dark matter is
estimated in different works to lie in the range of
\num{0.2}--\SI{0.6}{\giga\eV\,\cm^{-3}} \cite{c1_read}, and we adopt the value
of \SI{0.3}{\giga\eV\,\cm^{-3}} in our analysis.  In the galactic rest
frame, the velocity distribution in an isothermal halo is readily
described by a Maxwellian distribution:

\begin{equation}
    f(\textbf{v}) \mathrm{d}\textbf{v} = \frac{  \mathrm{d} \textbf{v}}
{(2 \pi)^{3/2}\sigma_\mathrm{rms,DM}^3} \exp \left( - \frac{|\textbf{v}|^2}
{2 \sigma_\mathrm{rms,DM}^2}\right)\,,
\label{eqn:velo_pro}
\end{equation}
where the velocity dispersion $\sigma_\mathrm{rms,DM} = \sqrt{3/2}\,
v_\mathrm{c}$ is related to the local circular rotational velocity,
$v_\mathrm{c}$, in the Galaxy \cite{turner1990}. Recent studies report local
galactic rotational velocities ranging from $(200 \pm 20)$\,\si{\kilo\meter\,\second^{-1}} to $(279 \pm 33)$\,\si{\kilo\meter\,\second^{-1}} \cite{c1_Green},
and we adopt a value of \SI{270}{\kilo\meter\,\second^{-1}}, following the analysis presented in
\cite{c1_kerr}.

In the laboratory frame, the velocity distribution,
$\textbf{v}_\mathrm{lab} = \textbf{v} - \textbf{v}_\mathrm{E}$, must
be corrected for the velocity of the Earth, $\textbf{v}_\mathrm{E}$,
with respect to the DM halo. The velocity $\textbf{v}_\mathrm{E}=
\textbf{v}_\mathrm{S} + \textbf{v}_\mathrm{O} + \textbf{v}_\mathrm{R}$
comprises contributions from the solar motion through the Galaxy,
$\textbf{v}_\mathrm{S}$, the Earth orbital velocity,
$\textbf{v}_\mathrm{O}$, and, in principle, also the velocity of Earth
rotation, $\textbf{v}_\mathrm{R}$. The $\textbf{v}_\mathrm{R}$ term
can be safely neglected, and the velocity distribution of dark matter
particles in the laboratory frame can be written in the following form
\cite{turner1990,ohare2017}:

\begin{equation}
f\,\mathrm{d} u = {\left[\frac{3}{2\pi} \right]^{1/2}} \frac{\mathrm{d} u}{r} 
{\exp[-3(r^2 + u)/2]} \sinh(3r\sqrt{u})\,,
\end{equation}
where $r = v_\mathrm{E}/\sqrt{2}\sigma_\mathrm{rms,DM}$, and the parameter $u$
reflects the relation between the DM particle mass $m_{\gamma'}$ and the
physical width of the distribution:

\begin{equation}
\Delta \nu = m_\gamma' v_{\gamma'}^2/2c^2 = 98\,\mathrm{Hz}\,(m_{\gamma'}\,c^2/\mu\mathrm{eV})\,u\,.
\end{equation}

The halo velocity distribution is focused at the lower end by the Sun capturing
DM particles with velocities smaller than the Sun escape velocity
\citep{griest1988}, $v_\mathrm{e,S} \approx 42.1$\,km/s ($u=0.16$). This effect
is small and can be ignored. At the upper end, the DM velocity distribution
should be truncated near the galactic escape velocity, $v_\mathrm{esc}$
\cite{c1_Undagoitia}. Numerical simulations indicate that the upper end
truncation occurs at velocities of 450--650\,km/s ($u_\mathrm{h} = 2.78$--5.80)
\cite{c1_Smith,pillepich2014}. 

\begin{figure}[h!]
\begin{center}
\includegraphics[width=0.95\textwidth]{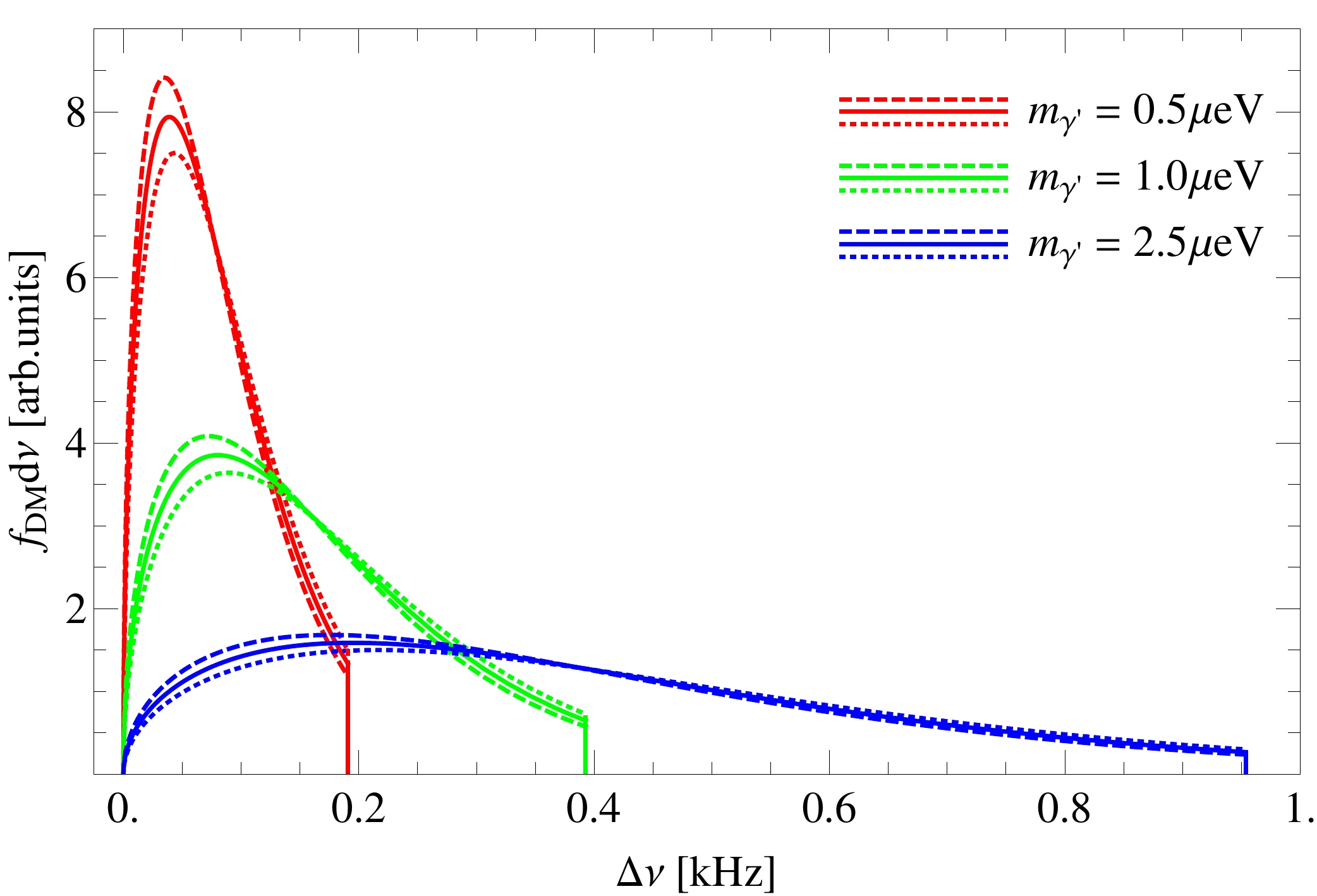}
\caption{Relative power and distribution of the dark matter
  conversion signal with respect to the rest-mass frequency at
  \SI{0.5}{\micro\eV} (\SI{122}{\mega\Hz}, red), \SI{1.0}{\micro\eV}
  (\SI{244}{\mega\Hz}, green), and \SI{2.5}{\micro\eV}
  (\SI{610}{\mega\Hz}, blue), calculated assuming the DM velocity
  truncation at $v_\mathrm{esc}=$\SI{530}{km.s^{-1}}. The dashed and
  dotted lines of respective colors illustrate the maximum effect of
  yearly modulation of the signal measured on December 2 (dashed) and
  June 2 (dotted).}
\label{fg:pdu}
\end{center}
\end{figure}

Recent analysis of the GAIA measurements \cite{deason2019} yields
$v_\mathrm{esc}=528^{+24}_{-25}\si{km.s^{-1}}$, and based on this
result, a truncation velocity of $v_\mathrm{esc}=\SI{530}{km.s^{-1}}$
is assumed throughout this paper.  The resulting expected shapes of
the dark matter signal are plotted in Fig.~\ref{fg:pdu} for particle
masses of 0.5, 1.0, and \SI{2.5}{\micro\eV}. For this range of mass,
the peak power changes roughly $\propto 1/m_{\gamma'}$.  The dark
matter signals also show a measurable effect of annual modulation due
to the orbital motion of the Earth. For the choice of the dark matter
halo parameters, the signal is modulated by about 12\% and shifted in
frequency by $\approx 16.4 (m_{\gamma'}/1\si{\micro\eV})\,\si{\Hz}$.

\begin{figure}[t!]
\begin{center}
\includegraphics[width=0.95\textwidth]{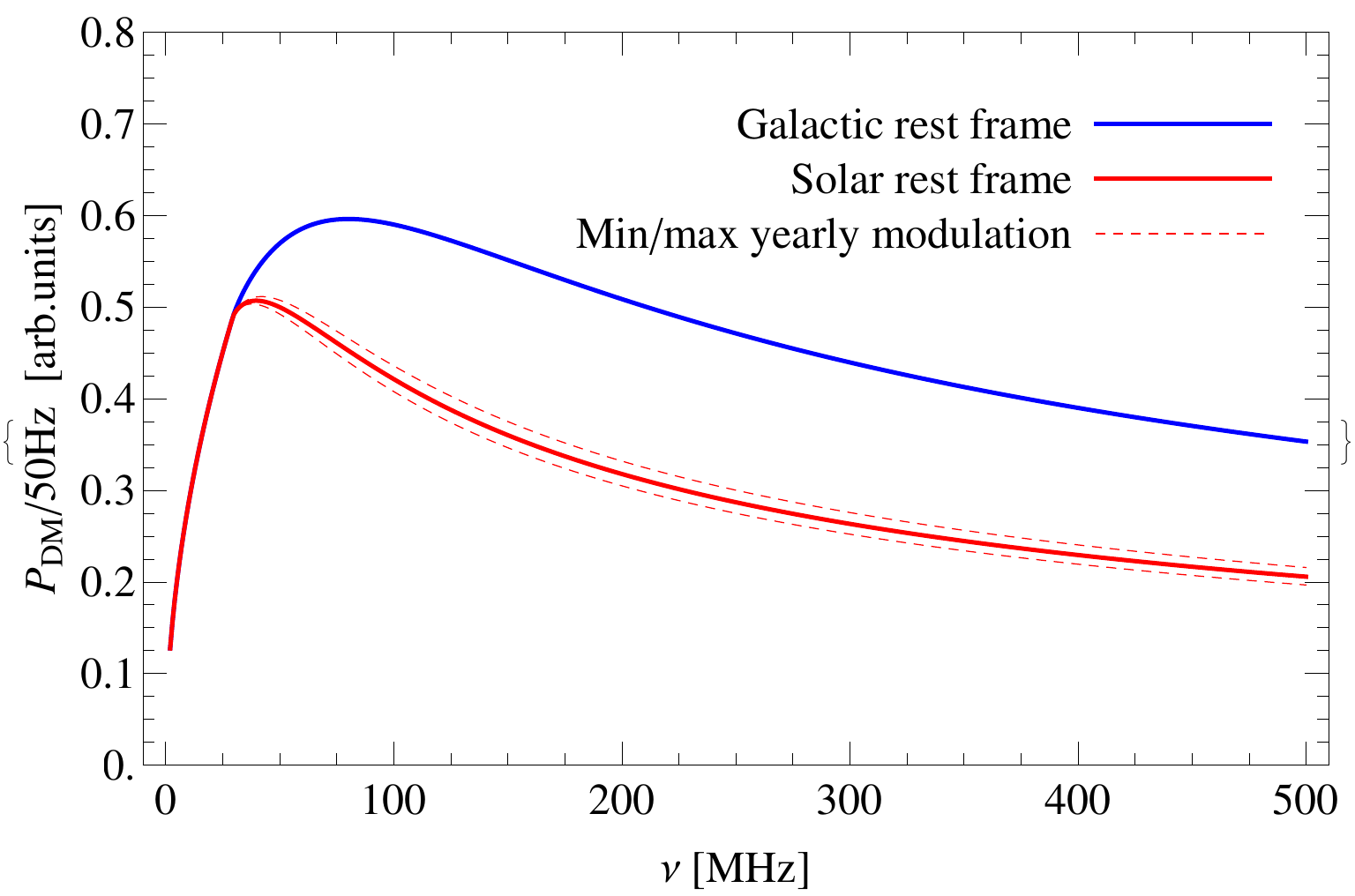}
\caption{Relative power of DM signal in a single \SI{50}{\Hz} channel of
  WISPDMX as a function of measurement frequency. The peak at
  \SI{39.8}{\mega\Hz} corresponds to the situation when the channel bandwidth
  of WISPDMX measurements roughly coincides with the bandwidth resulting from
  the velocity truncation of the DM halo. Above and below this frequency, the
  channel power is decreasing due to over- and under-resolving the DM
  signal. For the under-resolved signal at $\nu < \SI{39.8}{\mega\Hz}$, the channel
  signal is $\propto \nu^{1/2}$. For the over-resolved part, it depends on the
  actual shape of the DM velocity distribution.}
\label{fg:1ch}
\end{center}
\end{figure}

The relative power of the DM signal received in a single 50\,Hz
channel of WISPDMX, $P_\mathrm{DM}(\nu) =
P_\mathrm{DM}(\Delta\nu(\Delta u)=50\,\mathrm{Hz})$, is shown in
Fig.~\ref{fg:1ch} as a function of the measurement frequency.  The
power peaks at $\nu = \SI{39.8}{\mega\Hz}$, at which the channel width
of \SI{50}{\Hz} roughly corresponds to the bandwidth induced by the
velocity truncation in the dark matter halo. Below
\SI{39.8}{\mega\Hz}, the signal power decreases $\propto \nu^{1/2}$,
as expected for the case of under-resolving the signal. Above
\SI{39.8}{\mega\Hz}, the DM signal is over-resolved, and its
dependence on the measurement frequency is determined by the shape of
the DM velocity distribution.

\begin{figure}[ht]
  \begin{center}
    \includegraphics[width=0.95\textwidth]{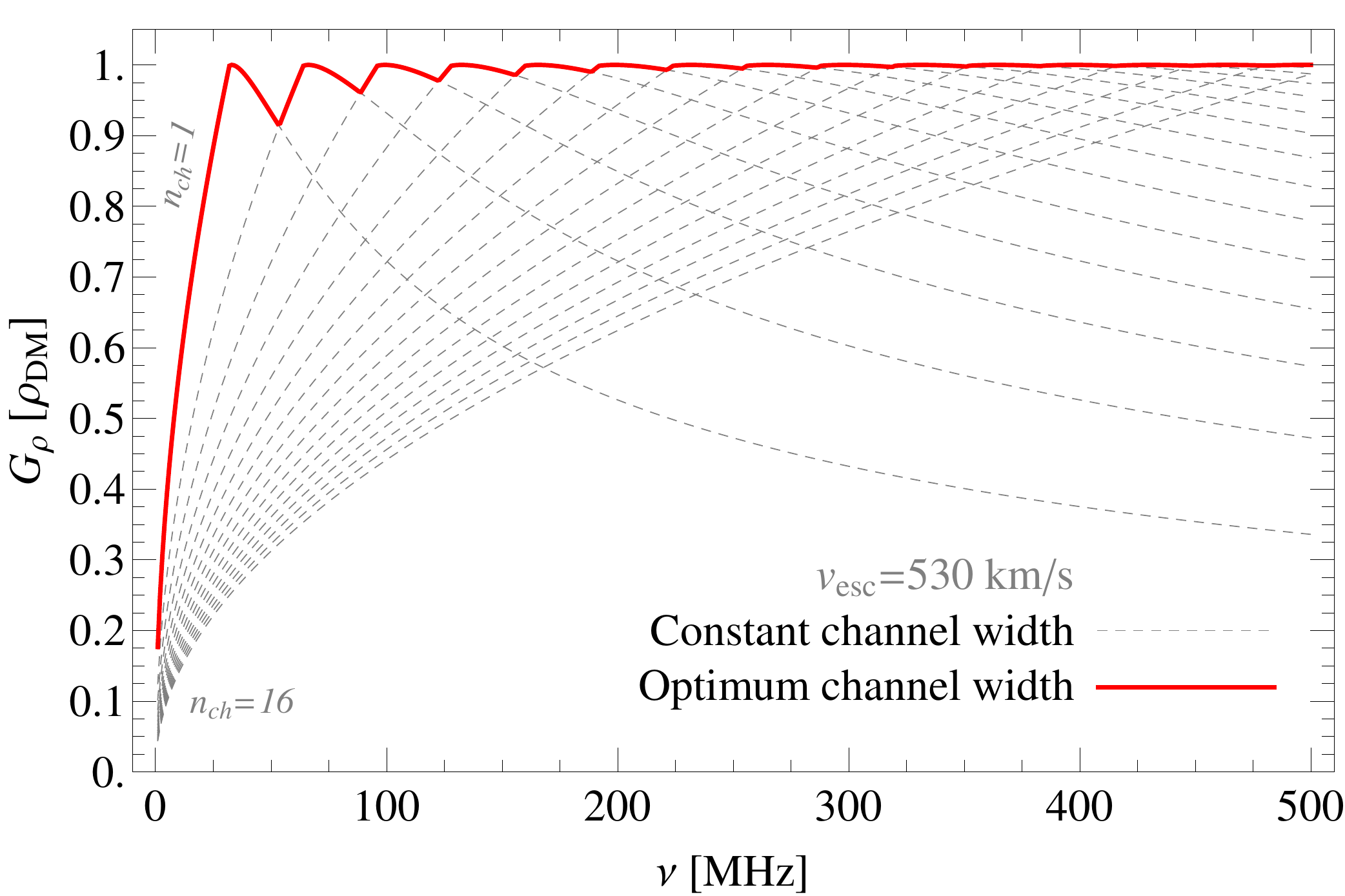}
    \caption{Sensitivity to the dark matter signal that can achieved
      with WISPDMX by summing over different numbers of
      \SI{50}{\hertz} spectral channels. The effective relative
      sensitivity factor $G_\rho$ is the ratio of the measurement
      sensitivity to the maximum sensitivity, calculated for
      $v_\mathrm{esc}=\SI{530}{\kilo\meter\,\second^{-1}}$. Dashed
      lines illustrate sensitivities resulting from summing over a
      fixed number of spectral channels, from one to sixteen. The
      upper envelope of these curves, plotted in red, describes the
      sensitivity of WISPDMX to the dark matter signal achieved by
      summing over the optimum numbers of spectral channels as listed
      in Table~\ref{tb:pixch}.}
    \label{fg:sumch}
  \end{center}
\end{figure}

For the over-resolved signal, summing over several frequency bins
should improve the sensitivity. The optimum sensitivity achieved with
the channel summing is illustrated in Fig.~\ref{fg:sumch} which shows
an effective relative sensitivity factor, $G_\rho$, for the dark
matter signal received in multiple channels of WISPDMX
measurements. In the calculations of the sensitivity of WISPDMX
measurements, this factor should be applied to modify the value of
local dark matter density. The different curves plotted in
Fig.~\ref{fg:sumch} illustrate that summing over up to 13 frequency
channel is required to achieve a nearly homogeneous sensitivity over
the entire \SIrange{40}{500}{\mega\Hz} frequency range.  The plots
shown in Fig.~\ref{fg:sumch} are calculated for $v_\mathrm{esc} =
530$\,km/s. The respective frequency ranges for achieving the optimal
sensitivity by summing over a given number of spectral channels are
given in Table~\ref{tb:pixch}.  The effect that the uncertainty in the
truncation velocity has on this optimum sensitivity curve is within
10\% for the plausible range of $v_\mathrm{esc}$ from 450\,km/s to
650\,km/s.

\begin{table}
  \caption{
  Frequency ranges for optimal sensitivity achieved by summing over $n_\mathrm{ch}$ 
  channels of 50\,Hz in width.}
  \label{tb:pixch}
  \begin{center}
    \begin{tabular}{r|rcr|r}\cline{1-2}\cline{4-5}
    \multicolumn{1}{c}{~} & & & \multicolumn{1}{c}{~} & \\[-2.5ex]\cline{1-2}\cline{4-5}
    \multicolumn{1}{c|}{$n_\mathrm{ch}$} & \multicolumn{1}{c}{Range [MHz]} & & 
    \multicolumn{1}{c|}{$n_\mathrm{ch}$} & \multicolumn{1}{c}{Range [MHz]} \\ \cline{1-2}\cline{4-5}
    1  &   0.0--~59.0 & &  7 & 263.8--304.8 \\ 
    2  &  59.0--100.6 & &  8 & 304.8--345.3 \\
    3  & 100.6--141.6 & &  9 & 345.3--386.4 \\
    4  & 141.6--182.2 & & 10 & 386.4--426.9 \\
    5  & 182.2--223.2 & & 11 & 426.9--468.0 \\   
    6  & 223.2--263.8 & & 12 & 468.0--500.0 \\ \cline{1-2}\cline{4-5}
    \end{tabular}
  \end{center}
\end{table}

\section{WISP Dark Matter eXperiment (WISPDMX)} \label{sec:onwispdmx}

\begin{figure}
    \centering 
    \includegraphics[scale =0.13]{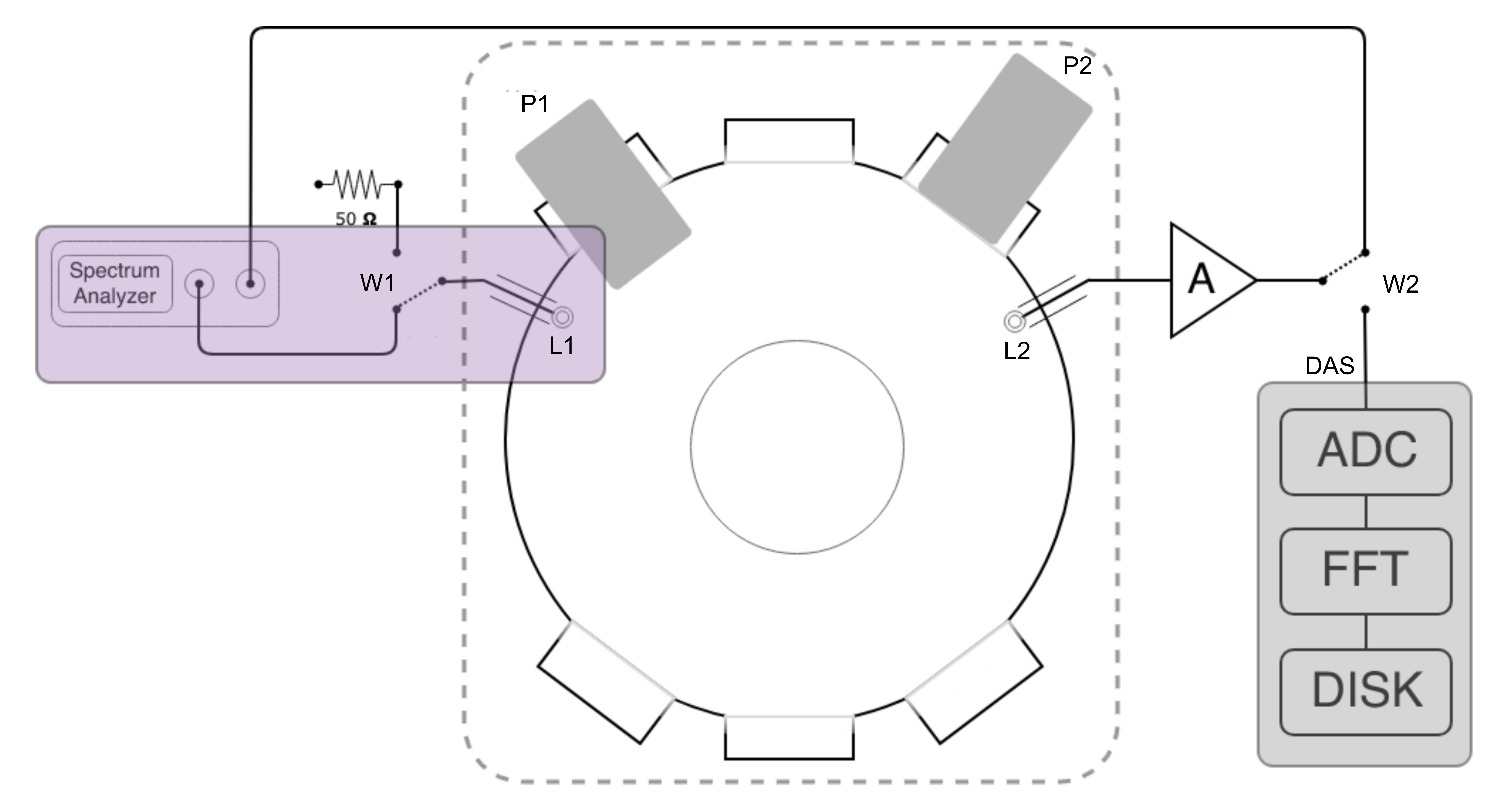}
    \caption{Schematic diagram of the WISPDMX setup, divided into three
      groups: \textbf{The Mechanical Components} (box with dash-line
      border) includes the 208 MHz RF cavity, and two tuning
      plungers $P1$ and $P2$. \textbf{The Frequency Calibration} (the blue
      shaded box) comprises a radio switch $W1$, a spectrum
      analyzer, and two loop antennas $L1$ and $L2$ coupled to the
      cavity. \textbf{The Acquisition System} (the gray shaded box)
      includes the antenna $L2$, amplifier chain ($A$) with an average
      gain of 80 dB, an analog-to-digital converter (ADC, Alazar
      ATS-9360), and a CUDA FFT unit (Nvidia GPU). The antenna $L2$ and
      the amplifier chain are shared between the calibration system
      and the acquisition system, using the radio switch $W2$.}
    \label{fig:wispdmx_setup}
  \end{figure}

The WISPDMX experiment employs a radio frequency resonant cavity of
the type used at the proton accelerator ring designed for the SPS
collider and modified for the HERA experiment
\cite{c2_cavity_3,c2_cavity_1,c2_cavity_2}.  A basic scheme of the
experiment setup is shown in Fig.~\ref{fig:wispdmx_setup}. The WISPDMX
apparatus comprises the following functional groups: mechanical
components (the cavity, the plunger assembly, and the driving motors),
an amplifier chain, a frequency calibration system, a data acquisition
system, and an automated experiment control system.  The cavity is
tuned with a set of two plungers which are driven by computer
controlled electric stepper motors.  A custom made data acquisition
system provides recording over the entire \SI{500}{\mega\Hz} frequency
range with a frequency resolution of \SI{50}{Hz}. The input voltage
signal is digitized in the time domain using a commercial 12-bit
analog-to-digital converter (ADC) of the type Alazar ATS~9360.  The
fast Fourier transformation (FFT) of the digitized output is
subsequently carried out with a CUDA\footnote{Compute Unified Device
  Architecture developed by Nvidia for parallel GPU programming.}
based parallel code running on a commercial GPU unit (GTX Titan X).
The following subsections will describe the function of each of these
groups, with specific details on the development and performance of
the individual components.

  \subsection{Mechanical Components}

\begin{figure}
\centering 
\includegraphics[scale =0.5]{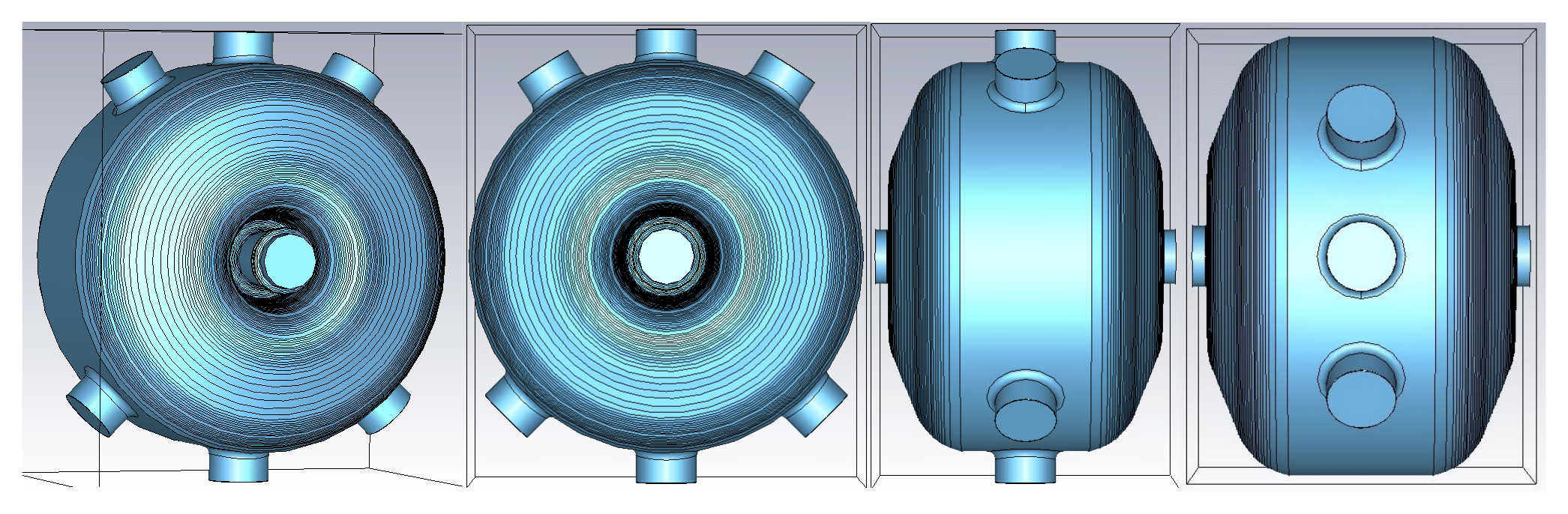}
\caption{A rendered view of the \SI{208}{\mega \Hz} resonant cavity at
  different orientations. This mechanical model is used with the
  commercial software package \textit{CST}\textsuperscript{TM}
  \cite{CST_simulation} to calculate the electromagnetic field
  configuration of each resonant mode.}
\label{fig:wispdmx_cst}
\end{figure}

The mechanical components of WISPDMX include a \SI{208}{\mega\Hz} HERA
resonant cavity and a tuning plunger assembly consisting of two
plungers as shown in Fig~\ref{fig:wispdmx_setup}. The cavity is
made of copper, and it has a diameter of \SI{96}{\cm} and volume of
\SI{447}{\liter} \cite{c2_cavity_1,c2_cavity_2,c2_cavity_3}
(Fig.~\ref{fig:wispdmx_cst}).  The inner surface of the cavity is
polished to maintain the electromagnetic reflection and achieve a
quality factor of $Q_0 \sim \num{e4}$ at a number of resonant modes.
The unloaded quality factor can be estimated from the following equation
\begin{equation}
Q_0 = \frac{\mu_0 V}{\mu_c S \delta_f}\,,
\end{equation}
where $\mu_0$ and $\mu_c$ describe permeability of vacuum and
copper respectively, $V$ and $S$ are the volume and total inner
surface of the cavity, and $\delta_f$ is the penetrating depth of an 
electromagnetic wave with a frequency $\nu$.

There are six ports distributed around the cylindrical rim of the
cavity. These ports initially used to mount tuning plungers and power
couplers when the cavity was operated as a part of particle accelerator. For
the WISPDMX setup, two ports are occupied by the tuning plungers, as
shown in Fig.~\ref{fig:wispdmx_setup}. The remaining ports are closed.
The positioning range of a plunger unit is constrained to be
\SIrange{0}{110}{\mm}. The stepper motor (Trinamic QSH6018) which can
be driven at a micro-step precision of 1/1600 of revolution is coupled
to a gear box and used for positioning the plunger in increments of
\SI{0.3}{\micro \meter} per micro step. The linear tuning distance of
\SI{1}{\mm} corresponds to two full revolutions of the gear box. Each
of the two stepper motors driving the plungers is monitored by the
device control computer which is synchronized with the data
acquisition control computer (see also
Sect.~\ref{acquisition_system}).

\subsection{Amplifier chain} \label{sec:amplifier}

\begin{figure}
\centering 
\includegraphics[scale =0.4]{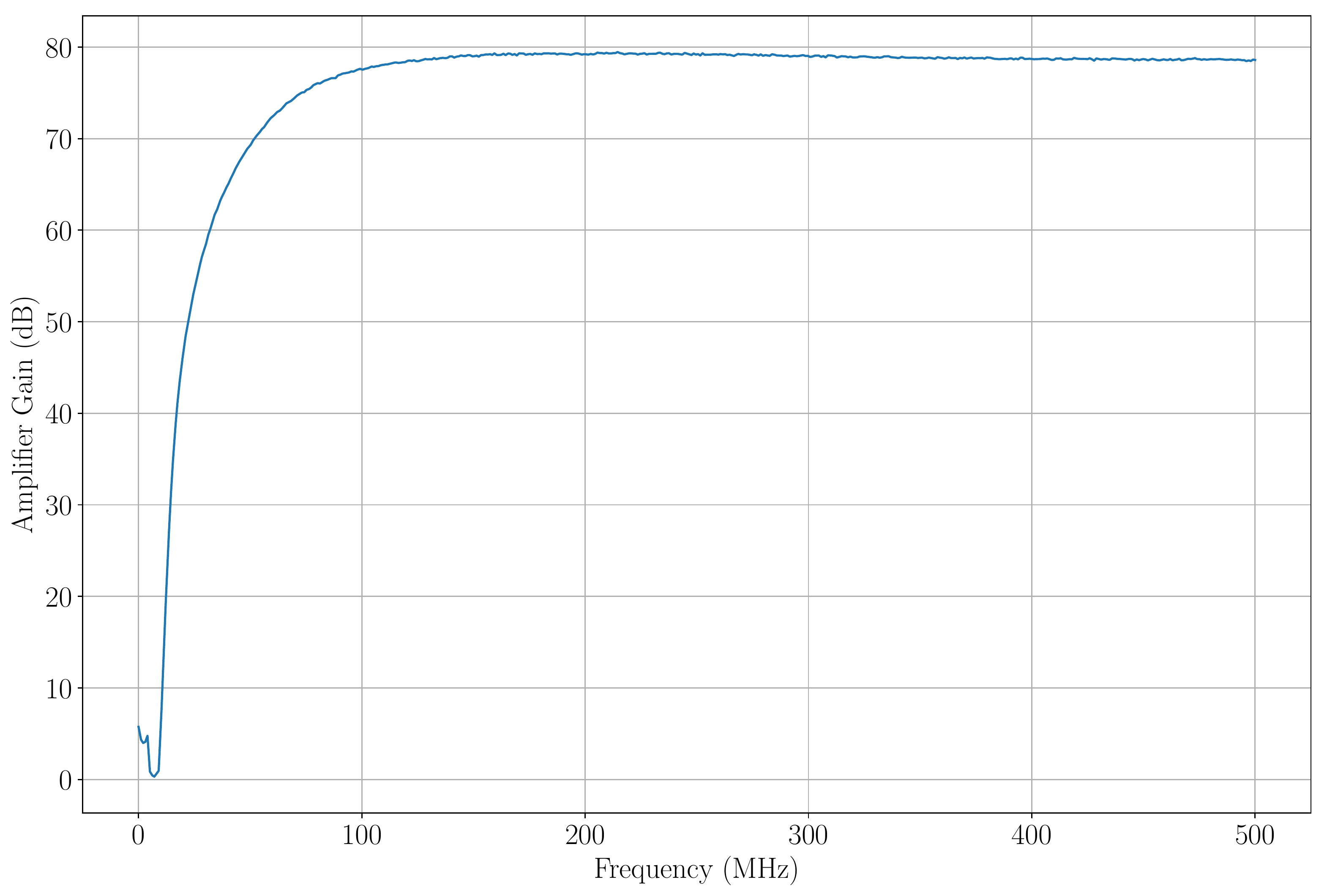}
\caption{Cumulative gain of the amplifier chain.  The
  nominal $\approx 79$\,dB amplification is realized over the
  \SIrange{150}{500}{\mega\hertz} frequency range, and the lowest
  usable frequency is $\approx 10$\,MHz ($\approx 0.4$\,neV).}
\label{fig:amp_factor}
\end{figure}

The amplifier chain of WISPDMX comprises two WantCom amplifiers,
WBA0105B and the WBA0105-45R, providing a total of 79\,dB
amplification over the \SIrange{150}{500}{\mega\hertz} frequency range
and an effective noise temperature of 35\,K. The cumulative gain
of the amplifier chain is shown in Fig.~\ref{fig:amp_factor}. Below
100\,MHz, the gain starts to decrease rapidly, but the amplifier chain
can be used effectively down to frequencies of $\approx 10$\,MHz, thus
extending the WISPDMX measurements down to the hidden photon mass of
$\approx 0.4$\,neV.

\subsection{The Resonant Cavity}
\label{section:simulation}

To describe and parameterize the performance of the HERA resonant
cavity over the entire \num{0}--\SI{500}{\mega\hertz} frequency range,
we combined transmission and reflection measurements and numerical
simulations of the cavity geometry. Within this range, we have
identified ten resonant modes of the cavity (see
Table~\ref{tab:10_modes}).  The individual modes have been
characterized using a transmission measurement carried out with a
spectrum analyzer Rohde-Schwarz FSP~7 outfitted with a tracking
function. These measurements provide the resonant frequency and the
quality factor of each of the resonant mode.

The form factor of each mode is obtained via equation
\ref{eqn:formfactor}, in which the electric field configuration of a given
resonant mode is calculated using a numerical simulation of the
cavity. The simulation software Computer Simulation
Technology\footnote{\textit{CST} Suite 2015 \& 2016 Version
  \cite{CST_simulation}} is used for calculating the field structure
of each resonant mode. The tunable plunger units are included in the
simulation as well.

In order to characterize the cavity response to the tuning plungers
and to determine the optimal tuning track, both the transmission
measurements and the field structure simulations have been made over a
12$\times$12 grid corresponding to all paired combinations of twelve
different tuning positions of each plunger equally spaced in the
\SIrange{0}{110}{\milli\meter} tuning range. At each individual step
of the simulation, the plunger positions are modified accordingly, and
the field configuration and the form factor are calculated for each of
the modes studied.

The resulting optimal tuning track is achieved by starting both
plungers at the zero position, then driving one of the plunger over
its full \SI{110}{\milli\meter} range, and then driving the second
plunger over the full tuning range (the actual choice of the order in
which the plungers are driven does not affect the tuning range and
speed, owing to the similarity in the plungers positions and
orientations with respect to the main axes of symmetry of the cavity).

The derived tuning response of the cavity is summarized in
Table~\ref{tab:10_modes}, in which ten resonant modes identified at
frequencies below \SI{500}{\mega\Hz} are listed together with their
calculated form factors and measured resonant frequencies.

Based on these calculations, we have selected four modes which have
non-zero form factors at all plunger positions and thus are best
suitable for our measurements.  The four selected modes are
$\text{TM}_{010}$, $\text{TE}_{111-1}$, $\text{TE}_{111-2}$, and
$\text{TM}_{020}$. 

\begin{figure}
\centering 
\includegraphics[scale=0.35]{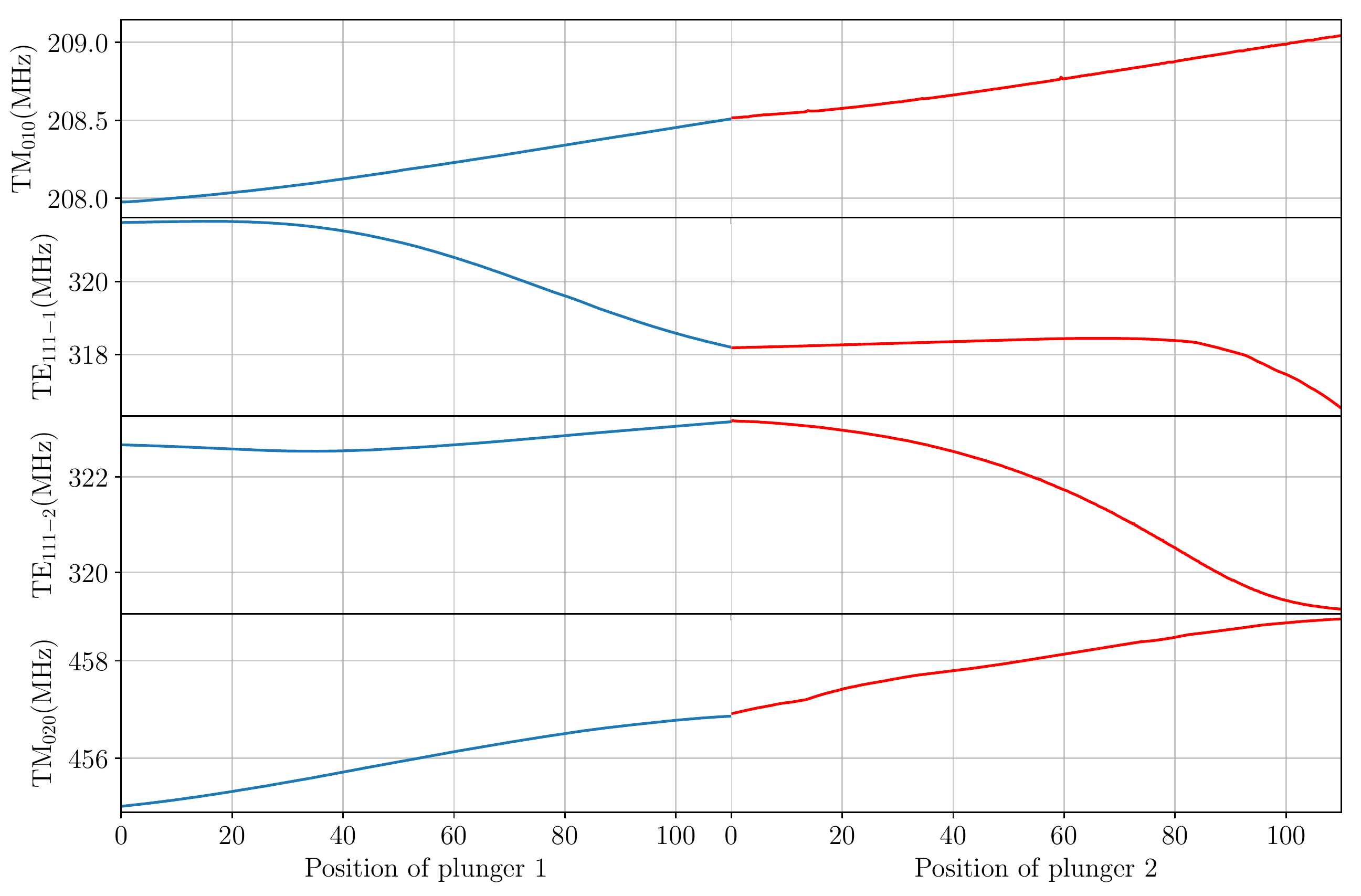}
\caption{Tuning ranges of the resonant frequencies of the four cavity
  modes corresponding to the tuning scheme adopted for the WISPDMX
  measurements. The two transverse magnetic modes, TM$_{010}$ and
  TM$_{020}$ are tuned at nearly constant rates of $\approx
  \SI{5}{kHz/mm}$ and $\approx \SI{20}{\kHz/mm}$, respectively. The
  two TE$_{111}$ modes are weakly tuned over most of the plunger
  ranges, and show the approximate highest tuning rates of
  $\SI{-45}{kHz/mm}$ (TE$_{111-1}$, for plunger 1 range of
  \SIrange{60}{100}{\milli\meter}) and $\SI{-60}{kHz/mm}$ (TE$_{111-2}$, for
  plunger 2 range of \SIrange{50}{90}{\milli\meter}).}
\label{fig:freqtuning}
\end{figure}

The measured changes of the resonant frequencies of these four
resonant modes for the full range of plunger positions used for the
tuning are shown in Fig.~\ref{fig:freqtuning}.  The plot shows that
the four resonant modes cover frequency ranges from $\Delta \nu\approx
\SI{1}{\mega\Hz}$ for the fundamental mode and up to $\Delta
\nu\approx\SI{6}{\mega\Hz}$ for the TE$_{111-2}$ mode. The combined
frequency range covered by the four modes is $\approx
\SI{10}{\mega\Hz}$. The ground mode is tuned at a rate of
$\approx \SI{5}{kHz/mm}$, over the full range of the plunger
tuning. The fastest tuning rate of about $\SI{-60}{kHz/mm}$ is
achieved for the TE$_{111-2}$ mode, in the range of
\SIrange{50}{90}{\milli\meter} of the second plunger. Thus, WISPDMX
can achieve an accuracy of frequency tuning of better than
\SIrange{1.5}{20}{\hertz} per tuning step, depending on the resonant mode.

Following these considerations, a tuning step of \SI{10}{\micro\meter}
is selected for the WISPDMX measurements, thus requiring a
total of \num{22000} steps to scan over the full range of the two
plungers. At this tuning step, the resonant modes are sampled at
frequency steps ranging from \SI{50}{\hertz} for the ground mode
TM$_{010}$ (corresponding to $\approx$1\% of its resonance width) to
\SI{610}{\hertz} (or $\approx$10\% of the resonance width) for the
fastest tuning rate of the TE$_{111-2}$ mode. Thus, the accuracy
WISPDMX measurements will not be affected by the frequency profiles of
the resonant modes.

The resulting frequency steps for both TM modes constitute 25\%
and 36\% of the respective DM signal width, hence the selected tuning
step will not affect the measurement sensitivity for these modes. The
same is true about the TE$_{111}$ mode, except for the ranges of its
high tuning rates where the tuning step in frequency reaches 100\% and
150\% (for TE$_{111-1}$ and TE$_{111-2}$, respectively) of the
expected width of the DM signal. However, even for these tuning steps,
the sensitivity reduction would be within 10\%. Hence the selected
plunger tuning step of \SI{10}{\micro\meter} is adequate for
scanning the frequency ranges covered by the four resonant modes used
in WISPDMX.

The tuning of the plungers changes both the
frequency as well as the field configuration inside the cavity.
Therefore, the tuning may lead to a change of form factor (see e.g.,
the form factor of the $\text{TM}_{011}$ where the deformation of the
field structure leads to an increase of the form factor when both
plungers are fully extending into the cavity). However, the form
factors of the four modes selected above vary only weakly with
changing the position of the plungers.  

It should be mentioned that both the simulations and the transmission
measurements have indicated that the peak frequencies of some of the
resonant modes have crossed in the course of the tuning over the full
range of the plungers.  However, none of the four modes selected for
making the WISPDMX measurements suffers from this mode crossing, as
can be seen from Fig.~\ref{fig:freqtuning}.

  \begin{table}
\caption{Characteristics of the HERA cavity at different plunger positions}
  \begin{center}
   \scalebox{0.8}{
\begin{tabular}{|l|ccc|ccc|ccc|} \hline
Mode: & \multicolumn{3}{c|}{Plunger position (0; 0)} & \multicolumn{3}{c|}{Plunger position (110; 0)} & \multicolumn{3}{c|}{Plunger position (110; 110)} \\
& $\nu_0$ & $Q_0$ & $\cal{G}$ & $\nu_0$ & $Q_0$ & $\cal{G}$   & $\nu_0$ & $Q_0$ & $\cal{G}$  \\ \hline
{\bf TM$_{010}$} & 207.99 & 53542 & 0.433  & 208.53 & 52395 & 0.431  & 209.06 & 51281 & 0.429  \\ 
TM$_{011}$ & 314.87 &       & 0.000        & 308.60 &       & 0.090       & 305.75 &       & 0.112        \\ 
{\bf TE$_{111-1}$} & 321.69 & 62067 & 0.679  & 318.13 & 55253 & 0.522  & 316.48 & 54482 & 0.504  \\ 
{\bf TE$_{111-2}$} & 322.69 & 62074 & 0.679  & 323.20 & 60431 & 0.677  & 319.21 & 54017 & 0.605  \\ 
TM$_{110-1}$ & 390.99 &       & 0.000        & 391.85 &       & 0.035        & 359.68 &       & 0.053        \\ 
TM$_{110-2}$ & 392.28 &       & 0.000       & 393.37 &       & 0.000        & 390.37 &       & 0.011        \\ 
TE$_{210-1}$ & 397.84 &       & 0.000        & 382.28 &       & 0.000       & 399.08 &       & 0.000        \\ 
TE$_{210-2}$ & 399.02 &       & 0.000        & 401.24 &       & 0.000        & 401.33 &       & 0.001        \\ 
{\bf TM$_{020}$} & 455.07 & 47902 & 0.321 & 456.92 & 46888 & 0.323  & 458.87 & 45739 & 0.324  \\ 
TM$_{012}$ & 461.73 &       & 0.000        & 457.33 &       & 0.018        & 449.94 &       & 0.009        \\ \hline
\end{tabular}}
\end{center} {\small{\bf Notes:}~Resonant frequencies, $\nu_0$, and
  geometrical form factors $\mathcal{G}$ of ten resonant modes of the
  HERA cavity identified in the \SI{500}{\mega\hertz} range. The mode
  parameters are evaluated at three different plunger positions: the
  initial $(0;\,0)$ position with both plungers fully retracted, the
  $(110;\, 0)$ position with one of the plungers extended by
  \SI{110}{\mm} into the cavity, and the $(110;\,110)$ position with
  both plungers fully extended into the cavity. Four modes with form
  factors close to unity (highlighted in boldface) are selected for
  the WISPDMX measurements. For these modes, the respective quality factors, $Q_0$, are
  also listed. }
\label{tab:10_modes}
\end{table}
 
\subsection{Calibration Procedures} 
\label{FC_sys} 

In order to monitor and calibrate the performance of the WISPDMX apparatus, several procedures are carried out at each tuning step of the measurement.

\subsubsection{Broadband gain}

\begin{figure}
\centering 
\includegraphics[scale =0.4]{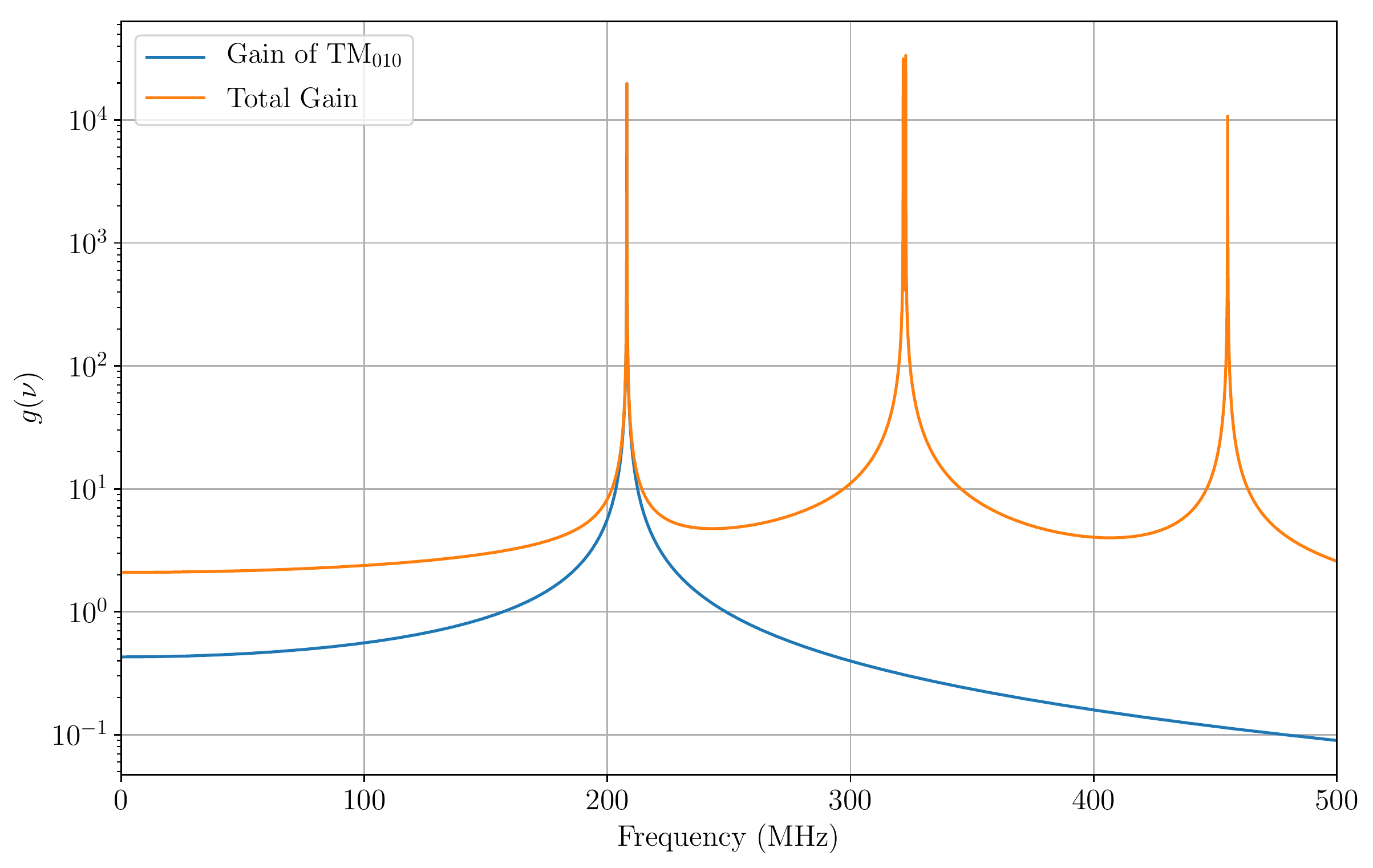}
\caption{The total
  cumulative gain $g(\nu)$ of the four relevant resonant modes of
  WISPDMX at the plunger assembly position (0,\,110), calculated using
  Eqn.~\ref{eqn:bbgain}. The contribution of the fundamental mode
  $\textrm{TM}_{010}$a is shown with a blue line.}
\label{fig:broadband_gain}
\end{figure}

With the four resonant modes identified in
Sect.~\ref{section:simulation} as suitable for hidden photon searches,
the broadband gain of WISPDMX can be calculated using
Eqn.~\ref{eqn:bbgain}. This calculation is performed at each tuning
step of the measurements. An example of the broadband gain calculated
for a single position of the plunger assembly is shown in
Fig.~\ref{fig:broadband_gain}.

\subsubsection{Monitoring of the resonances}

\begin{figure}
\centering 
\includegraphics[scale =0.3]{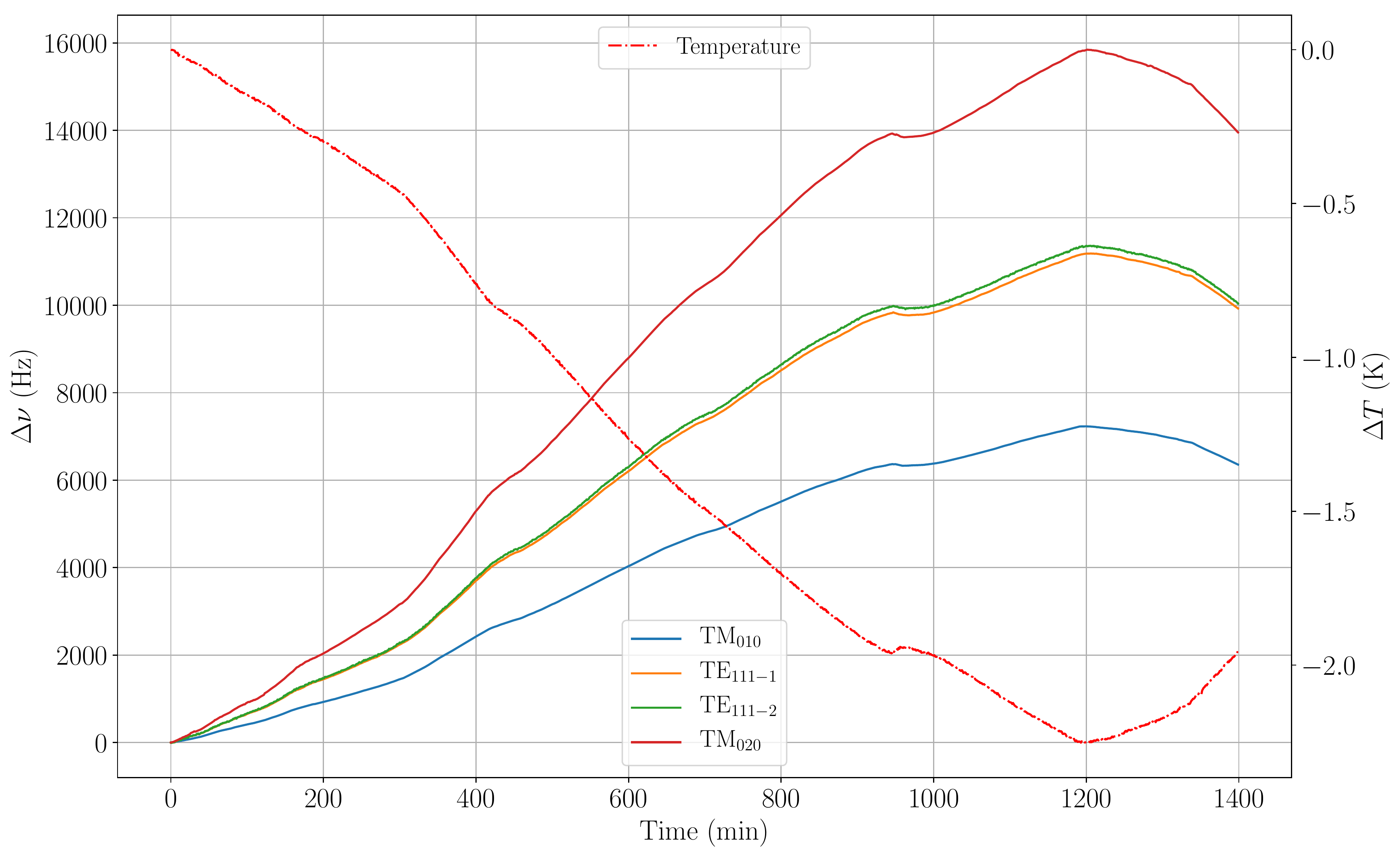}
\caption{Change of the resonant frequencies of the four selected modes
  induced by the decrease and increase of the ambient temperature. The
  change of the cavity volume with decreasing/increasing room
  temperature results in reciprocal changes of the frequencies of the
  resonant modes.}
\label{fig:Temperature_shift}
\end{figure}

In addition to the changes of the resonant response of the cavity
resulting from the tuning, variable ambient conditions (temperature,
air pressure) lead to variations of the resonant frequencies (the
changes of the respective quality factors and form factors are
negligible).  In Fig.~\ref{fig:Temperature_shift}, we show a typical
day-night cycle in the laboratory during which the temperature drops
by about $\Delta T\approx\SI{2}{\degreeCelsius}$.  This leads to an
increase of the resonant frequencies of the four modes by $\Delta
\nu(\mathrm{TM}_{010}) \approx \SI{7}{\kilo\Hz}$ up to $\Delta
\nu(\mathrm{TM}_{020})\approx \SI{16}{\kilo\Hz}$ with $\Delta
\nu_i/\nu_i\propto -\Delta T/T$, as expected for a self-similar
thermal contraction of the cavity.

A frequency calibration system is therefore necessary in order to
monitor the resonant frequencies during the experimental
measurements. The frequency calibration measurements are performed
intermittently (at intervals short enough to warrant interpolation of
the resonant frequencies and quality factors), each time after a
series of data taking and tuning steps. The calibration procedure uses
measurements of the transmission coefficient of the network which
contains the resonant cavity. The transmission coefficient is measured
by sending a feed signal and recording the output with the spectrum
analyzer.

The spectrum analyzer (RS FSP7 with tracking option) is connected to
two loop antennas ($L1$ and $L2$, in Fig.~\ref{fig:wispdmx_setup}), so
that the tracking generator feeds the antenna $L1$ and the
spectrum analyzer receives the feedback output signal from the
antenna $L2$.  The output from $L2$ can be switched by the radio
switch $W2$ between the input of the spectrum analyzer or the ADC of
the acquisition system.  The tracking generator connected to the antenna $L1$
sweeps the preselected frequency interval of \SI{300}{\kilo\Hz} around 
the resonant frequency of a given mode, sending a signal with a
power of \SI{-10}{\deci\belm} and a spectral resolution of $\Delta
\nu = \SI{300}{\Hz}$.

\begin{figure}
\centering 
\includegraphics[scale =0.3]{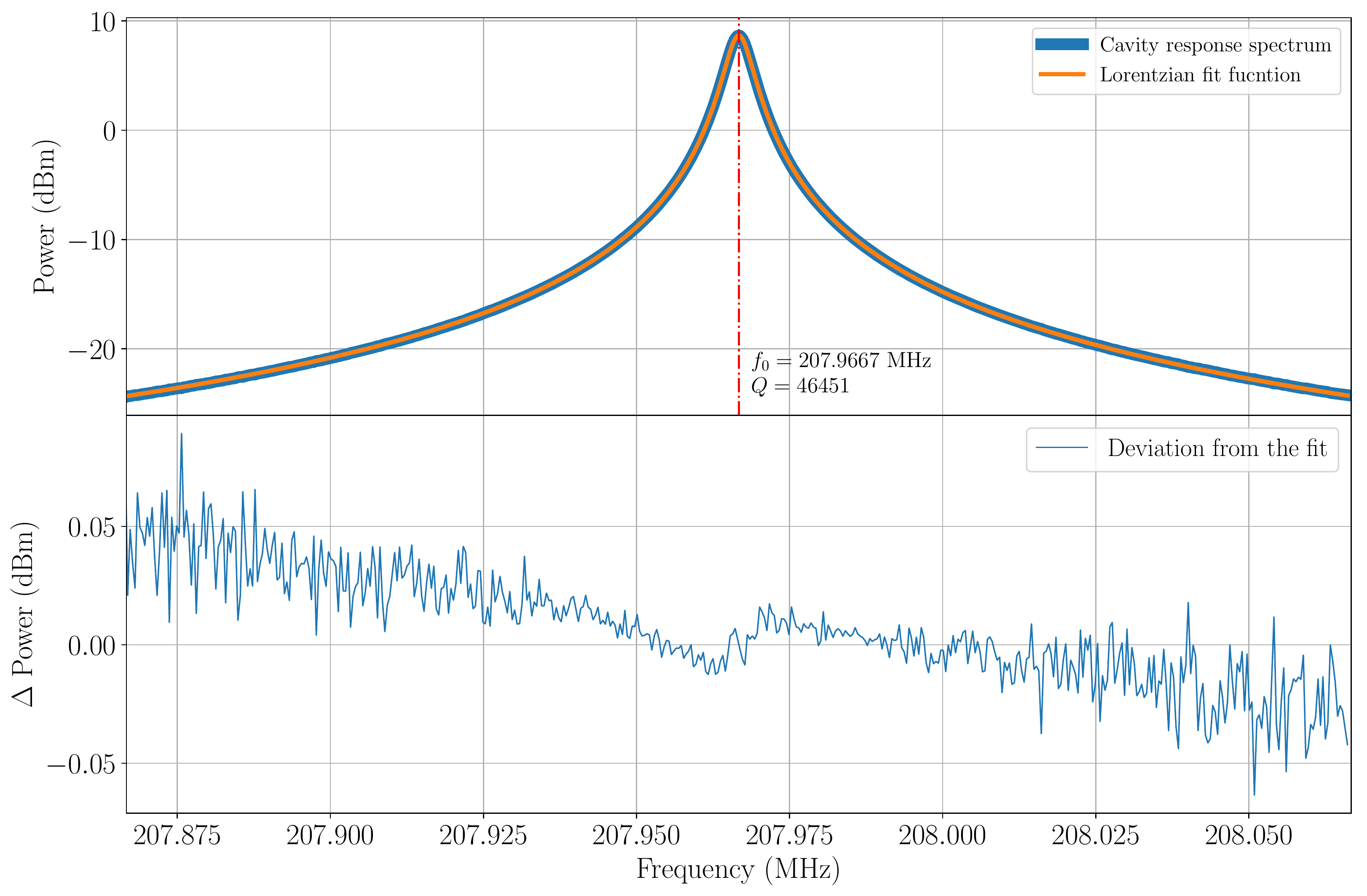}
\caption{Example of a Lorentzian profile fit to the transmitted power
  recorded near the resonant frequency of the $\text{TM}_{010}$ mode.
  The bottom panel shows the post-fit residual power.}
\label{fig:Lorentzian_fit}
\end{figure}

From the output registered at the antenna $L2$, the central resonant
frequency and the loaded quality factor are determined by fitting a
characteristic Lorentzian function given by
    
\begin{equation}
P_\mathrm{res} =\frac{ h }{1+4Q^2(\frac{\nu}{\nu_0}-1)^2 },  
\label{eqn:Lorentzian_func}
\end{equation}
where $P_\mathrm{res}$ is the power recorded by the spectrum analyzer,
$\nu_0$ and $Q$ are the resonant frequency and quality factor,
respectively.  The scaling parameter $h$ is proportional to the
strength of the input signal.  A least square fit is used to estimate
the values of these three parameters and their uncertainties. The
fit yields a good agreement with the measurement as illustrated in
Fig.~\ref{fig:Lorentzian_fit}.

\subsubsection{Reference frequency}

The overall accuracy of the frequency calibration may depend also
on the systematic uncertainty and drifts of the reference frequency of
the spectrum analyzer. The reference frequency can drift over time
and with changing temperature conditions.  The temperature in the
laboratory changes by as much as \SI{2}{\degreeCelsius}. The spectrum
analyzer was calibrated by the manufacturer in December 2016, and
the frequency measured with the spectrum analyzer has a systematic
uncertainty at the level of \SI{e2}{\Hz}, taking into account that for
the Rohde-Schwarz FSP 7, the relative aging rate
$\dot{\nu}=\SI{e-6}{\per\year}$ and the relative frequency drift per
\si{\degreeCelsius} is \num{e-6}.  In addition to the systematic
uncertainty, we estimate the statistical uncertainty from the fitting
of the Lorentzian function \ref{eqn:Lorentzian_func} to be lower than
$\SI{1}{\Hz}$ (see also Fig.~\ref{fig:Lorentzian_fit}). As all
these factors are substantially smaller than the intrinsic systematic
uncertainty of the reference frequency, we take $\sigma_\nu =
\SI{e2}{\Hz}$ as the total frequency accuracy of our
measurements. This corresponds to an error of $\sigma_\mathrm{m} =
\SI{0.4}{\pico\eV}$ for the hidden photon mass, and hence this
uncertainty would not affect the present WISPDMX measurements. It may
however become a potentially adverse issue if one would attempt to
increase the signal-to-noise ratio of the resonant detection by
summing multiple independent measurements.

\subsubsection{Antenna coupling}

\begin{figure}
\centering 
\includegraphics[scale =0.35]{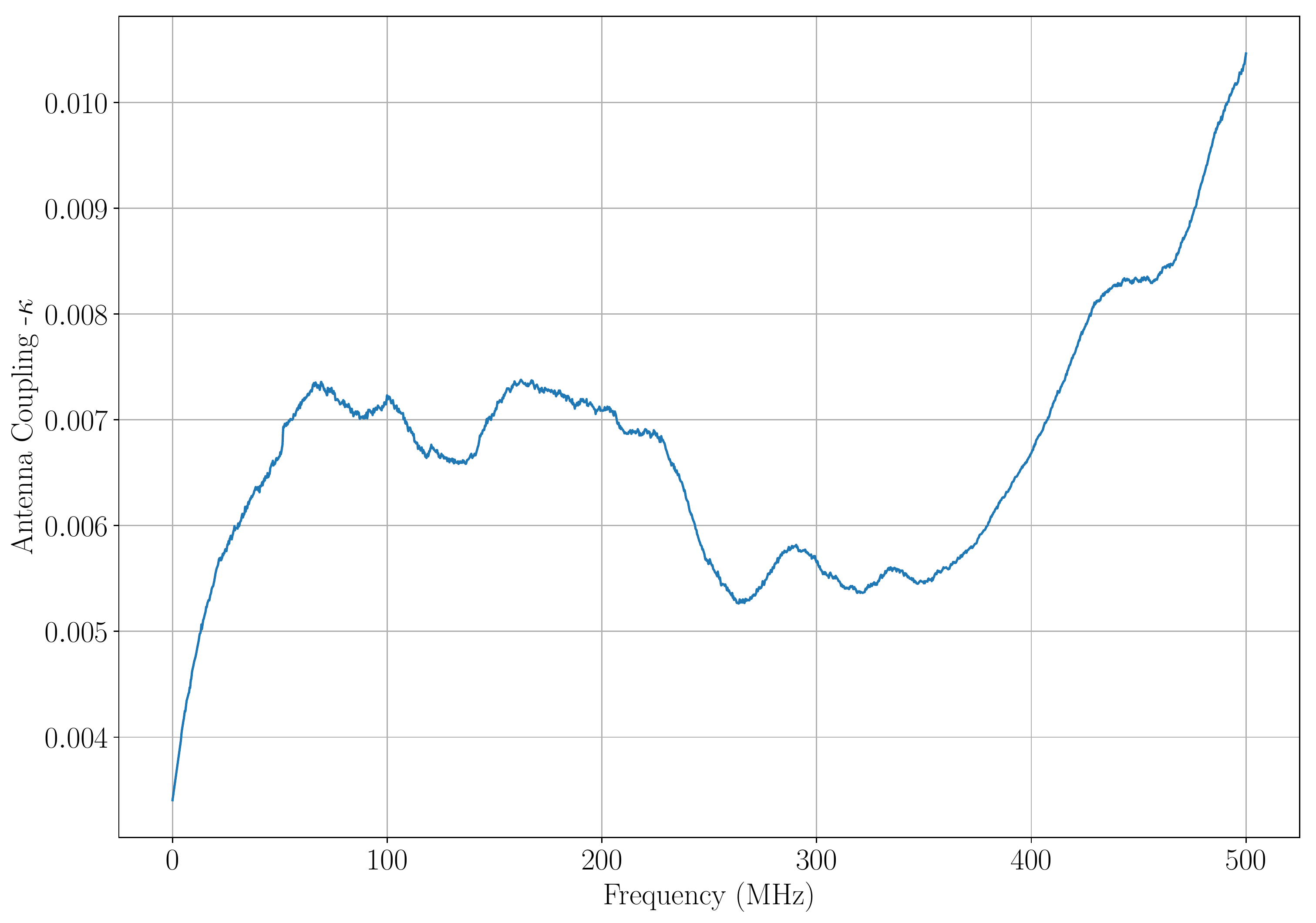}
\caption{The coupling factor, $\kappa(\nu)$, of the antenna used for the WISPDMX measurements.}
\label{fig:antenna_coupling}
\end{figure}

A network analyzer, Anritsu 37369A, is connected to the loop
antenna $L2$ to measure the coupling, $\kappa$, of the antenna to
the cavity over the entire measured
bandwidth. The coupling is derived using the $S_{11}$ reflection parameter, with
\begin{equation}
\kappa = \frac{1- |S_{11}|}{1+|S_{11}|}\,.
\label{eq:kappa}
\end{equation} 
The measured antenna coupling of the WISPDMX setup is shown in
Fig.~\ref{fig:antenna_coupling}. The WISPDMX antenna is weakly coupled
to the cavity, with $\kappa\ll 1$ over the entire bandwidth. Our
measurements indicate that the coupling is not sensitive to the cavity
tuning, and it varies within less than 5\% over the entire range of
the plunger positions. We apply therefore a single $\kappa(\nu)$ profile for
evaluating all of the WISPDMX measurements.

\subsection{Data Acquisition System} \label{acquisition_system}

The data acquisition system (DAS) of WISPDMX provides real time
recording  at a rate of
\SI{1}{\giga \Sample \per \second}, corresponding to a maximum recording bandwidth
of \SI{500}{\mega\hertz}. The system contains two main
components: a 12-bit analog-to-digital converter (ADC), modified for
performing a high volume continuous streaming to the host memory,
and a high-power CUDA GPU employed for carrying out the FFT and
related operation (e.g., array adding and converting).

\begin{figure}
\centering 
\includegraphics[scale =0.45]{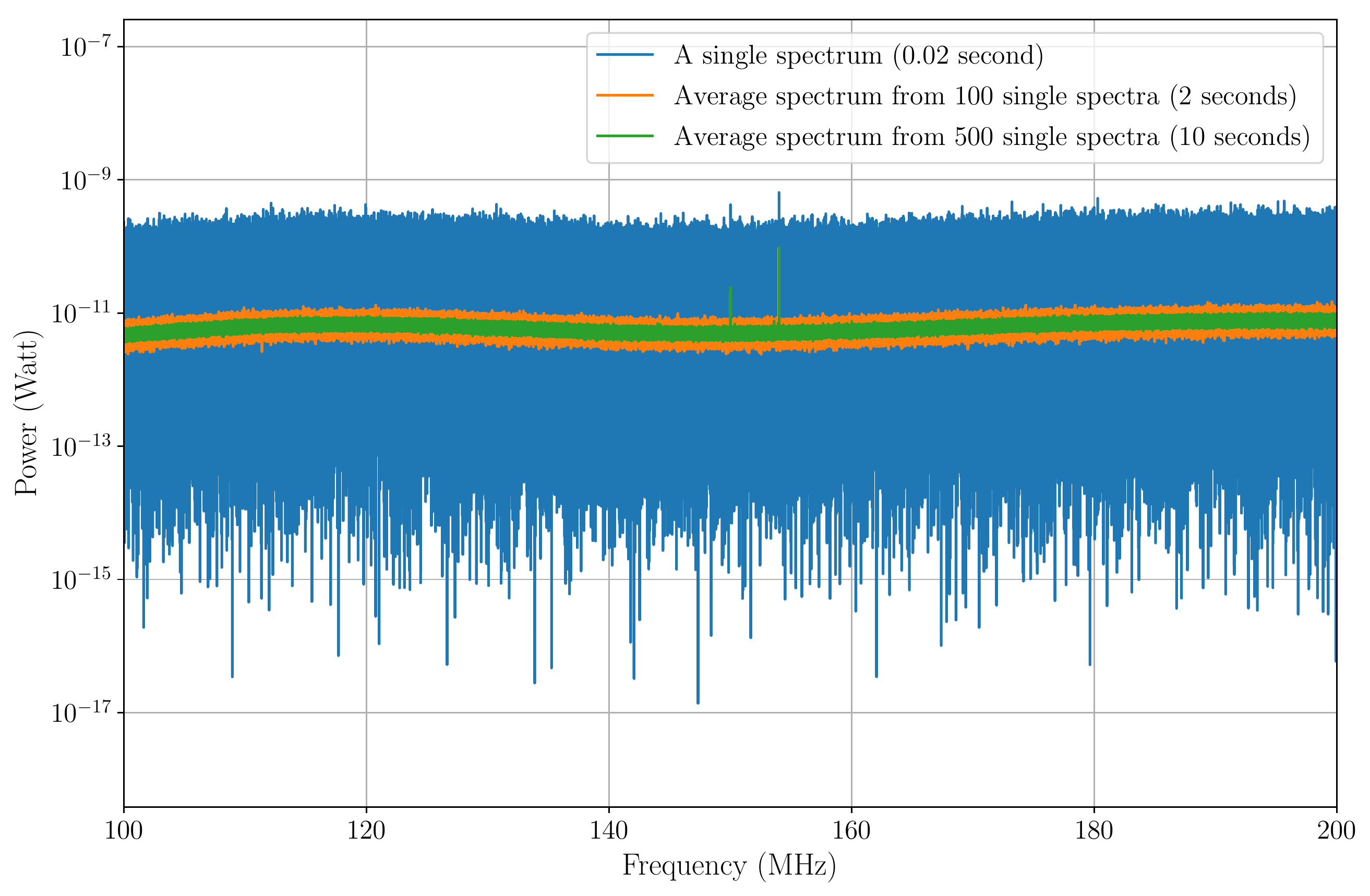}
\caption{Examples of broadband spectra from the acquisition system
  illustrating the noise reduction achieved by spectral
  averaging. \textbf{Blue}: single spectrum from \SI{0.02}{\second} of
  data-taking. \textbf{Green}: average of 100 single spectra.
 \textbf{Red}: average of 500 single spectra. The latter
  is the output spectrum for each acquisition step of WISPDMX.}
\label{fig:spectrumaveraging}
\end{figure}

The DAS can presently accumulate up to \SI{10}{\second} of
continuous sampling in the time domain, hence it can principally
achieve a spectral resolution of \SI{0.1}{\Hz} over the entire
\SI{500}{\mega\Hz} band.  However, storing a high-resolution broadband
spectrum leads to prohibitively long dead times for the pipeline
processing, data I/O and storage. As a compromise between increasing
the spectral resolution and reducing the dead time, the
\SI{10}{\second} of data are broken down into 500 segments of
\SI{0.02}{\second} in duration, and the FFT is applied to each of
these segments. The resulting broadband spectrum has then a spectral
resolution of \SI{50}{\Hz}.  After the FFT, the resulting 500 spectra
are co-added to calculate the average spectrum.  After the averaging,
the noise in the averaged spectrum is $\sigma_\text{a} =
\sigma_1/\sqrt{N}$, where $\sigma_1$ is the noise power of a single spectrum
and $N$ is the number of spectra used in the averaging process.
The noise improvement achieved by the averaging of the individual spectra
is illustrated in Fig.~\ref{fig:spectrumaveraging}.

\begin{figure}
\centering 
\includegraphics[scale =0.45]{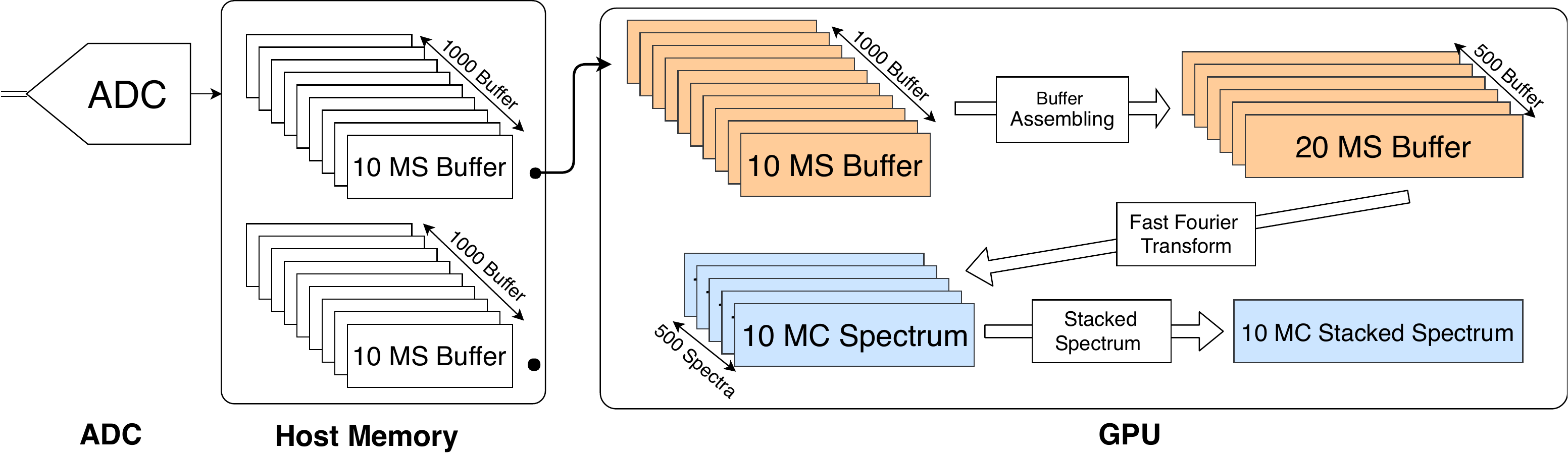}
\caption{Detail of the DAS operations, illustrating the
  multi-threading and memory management employed to process individual
  10-megasamples (10\,MS) time series recordings into 10-megachannels
  (10\,MC) power spectra. The GPU executes three main CUDA-powered
  functions: the buffer assembling, the CUDA-FFT, and the spectrum
  stacking.}
\label{fig:acquisition_diagram}
\end{figure}

A scheme of the WISPDMX data acquisition and processing procedures is
presented in Fig.~\ref{fig:acquisition_diagram}.  At each data
acquisition step, the ADC streams the data directly into the host
memory grid, splitting the data stream into \num{1000} individual
buffers each holding \SI{10}{\mega\Sample} (\SI{0.01}{\second}) of
data.  Once the first \num{1000} buffers are stored in the host
computer, the ADC starts recording the next \SI{10}{\second}-segment
of data and streams it to the second memory grid.  In parallel to the
data taking, the recordings from the first memory grid are transferred
to the GPU. Inside the GPU, each two consecutive \SI{10}{\mega\Sample}
buffers are merged to create \num{500} buffers containing
\SI{20}{\mega\Sample} time series arrays.  Each individual
\SI{20}{\mega\Sample} array is then Fourier transformed, yielding a
power spectrum with 10 million spectral channels (10\,MC spectrum, in
Fig.~\ref{fig:acquisition_diagram}) with a \SI{50}{\hertz} frequency
resolution. The resulting 10\,MC power spectrum is moved to the second memory
grid.

After all 500 spectra have been transferred to the second memory grid,
the average spectrum is calculated as the final output of the DAS for
a \SI{10}{\second} integration time corresponding to a single step of
the WISPDMX measurement run.

With the application of multi-threading and memory managing, the
computational processes inside the GPU are completed in less than
\SI{3.2}{\second} in total, (see the time breakdown in Table
\ref{tab:time_breakdown}). During this time, a tuning step
($\approx$\SI{0.1}{\second}) and the ADC initialization for the next
acquisition ($\le$\SI{2.8}{\second}) are performed in parallel. The
combined ADC and GPU acquisition system has therefore a fractional
dead time of less than \SI{30}{\percent}. In order to improve this
performance and achieve near real-time acquisition operations, it
would be required both to use multiple GPUs and to reduce the
initialization time of the ADC before each data acquisition. However,
this initialization time is a design limitation of the commercial ADC broad. 
 
    \begin{table}
      \caption{Chronological breakdown of a single data acquisition step}
      \begin{center}
      \begin{tabular}{c|c|c}
      \hline
      \hline
      \textbf{Component}            & \textbf{Process}                                                                       & \textbf{Time (\si{\second})} \\ \hline
      \hline
      \multirow{2}{*}{ADC} & Initialization of acquisition                                                                & 1.2 -- 2.8         \\ \cline{2-3} 
                                    & Filling host memory                                                                     & 10.002            \\ \hline
      \multirow{5}{*}{GPU}          & \begin{tabular}[c]{@{}c@{}}Data transfer from \\ host memory to GPU\end{tabular} & 2.0509            \\ \cline{2-3} 
                                    & Buffer assembling                                                                      & 0.21              \\ \cline{2-3} 
                                    & CUDA-FFT                                                                                 & 0.841             \\ \cline{2-3} 
                                    & Spectrum stacking                                                                      & 0.163             \\ \cline{2-3} 
                                    & Copying stacked spectrum to host                                                          & 0.008             \\ \hline
      Host                          & Saving stacked spectrum to disk                                                                           & 0.04              \\ \hline
      \end{tabular}
      \end{center}
{\small{\bf Note:}~The 
        processes in the GPU and the Host are run in parallel with the ADC 
        processes, which enables performing near real-time data acquisition.}
      \label{tab:time_breakdown}
    \end{table}

\subsection{WISPDMX data taking and calibration procedure}

\begin{figure}
  \centering 
  \includegraphics[scale =0.85]{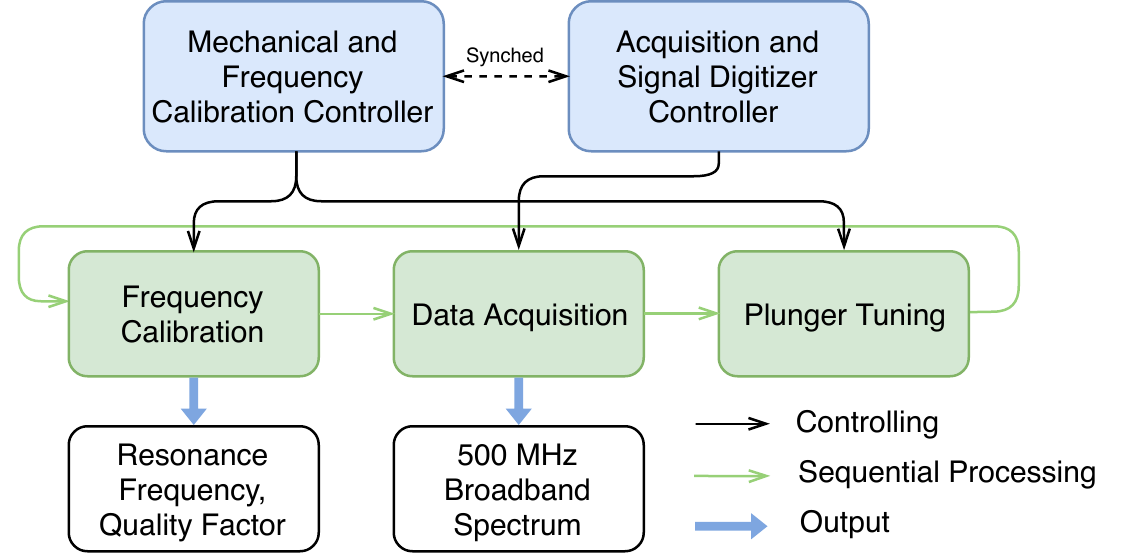}
  \caption{Operational scheme of WISPDMX measurements. The
    \textit{Control Module} (top row) drives the \textit{Executive
      Module} (middle row) which produces the science and calibration
    output data (bottom row). The \textit{Control Module} synchronizes
    the operations of the mechanical, frequency calibration, and data
    acquisition controllers. A measurement cycle begins with engaging
    the \textit{Frequency Calibration} block which determines the
    actual parameters of the resonant modes. The \textit{Data
      Acquisition} block is then initialized and set to record 10
    seconds of data. After the recording has been completed, the
    \textit{Control Module} first engages the \text{Plunger Tuning}
    block to move the plunger to a new tuning position (while
    transferring the recorded data to the GPU for further processing)
    and then proceeds to performing the next round of data
    acquisition. The data acquisition and plunger tuning steps are
    repeated ten times before starting the next measurement cycle
    which begins with execution of another frequency calibration
    procedure.}
  \label{fig:flowwispdmx}
\end{figure}

In order to make automated measurements at each of the \num{22000}
tuning steps planned for WISPDMX data taking run, the three main
components of WISPDMX described above are combined in a connected and
synchronized setup controlled by two dedicated computers: the
mechanical and frequency calibration controller managing the plunger
tuning, frequency calibration system, and radio switches, and the
acquisition and signal digitizer controller operating the DAS. A scheme
of the operational structure of WISPDMX measurements is shown in
Fig.~\ref{fig:flowwispdmx}.

The WISPDMX operations are ultimately structured into multiple
measurement cycles. Each measurement cycle begins with a frequency
calibration step (lasting for about \SI{4}{\second}). Subsequently,
the data acquisition system acquires data for the current setting of
the plunger assembly, and then the plunger assembly is moved to the
next tuning position. These two operations are repeated ten times,
before proceeding to the next measurement cycle. The output of both
the frequency calibration module and the data acquisition system is
stored on a local hard drive for offline processing and analysis.  The
respective dead time for a single WISPDMX measurement cycle is less
than \SI{30}{\percent}.

The typical total duration of a single measurement cycle is
$\le$\SI{133}{\second}. 
Within this time, the potential frequency drifts due to the
temperature changes should be $\le$\SI{40}{\hertz}, for the worst
measured temperature gradient ($\approx$\SI{0.14}{K/hr}, see
Fig.~\ref{fig:Temperature_shift}) and the most temperature-sensitive TE$_{111}$
mode. Thus even the maximum expected frequency drift should be smaller than the
frequency resolution of the measurements, and it can be adequately represented
by a linear
interpolation between two successive frequency calibration steps.

With the tuning step of \SI{10}{\micro \meter}, the accumulated shift
between two runs of the frequency calibration amounts to
\SI{100}{\micro \meter}.  This corresponds to a maximum possible
frequency change of \SI{6}{\kilo\hertz}, within which the frequency
changes are well described by a linear interpolation. The same
considerations apply to the quality factors and mode coupling factors
of the modes. Hence, the parameters of the resonant modes are
linearly interpolated, in order to obtain the calibration information
for each of the tuning steps performed between these two calibration
measurements.

\section{WISPDMX First Science Run}
\label{sec:analysis}

The measurements comprising the first science run of WISPDMX were
made during the time period from 23rd October 2017 to 2nd November
2017, comprising a total of \num{22000} spectra, each
produced from a single \SI{10}{\second} data taking and tuning step.
The first science run of WISPDMX had been carried out using a weakly
coupled antenna.

In the course of the measurements, the two plungers were consecutively
extended into the cavity, advancing at a rate of \SI{10}{\um} per
tuning step.  The complete run has been accomplished within \num{8}
sub-runs, with each sub-run providing \num{2000} or \num{3000}
acquisitions. The total acquisition time ($\num{22000} \times
\SI{10}{\second}$) corresponds to \SI{2.46}{\day} of data.  The
remaining time ($\sim 7$ days) comprised various auxiliary processes
and activities, including a remaking of one sub-run because of a
memory jam in the ADC, frequency calibration, plunger tuning, and
electronic switch operation, and a break between two sub-runs. For the
first science run, WISPDMX was operated at the room temperature of
\SI[separate-uncertainty=true]{20(1)}{\degreeCelsius}.  The frequency
calibration procedure was invoked after every 10 consecutive data
recording and tuning steps.  For the offline analysis, the resonant
frequencies and quality factors of the resonant modes are extracted
from the sweep spectra provided by the frequency calibration
procedure.

\subsection{Single Acquisition Spectra}

\begin{figure}
\centering 
\includegraphics[scale =0.35]{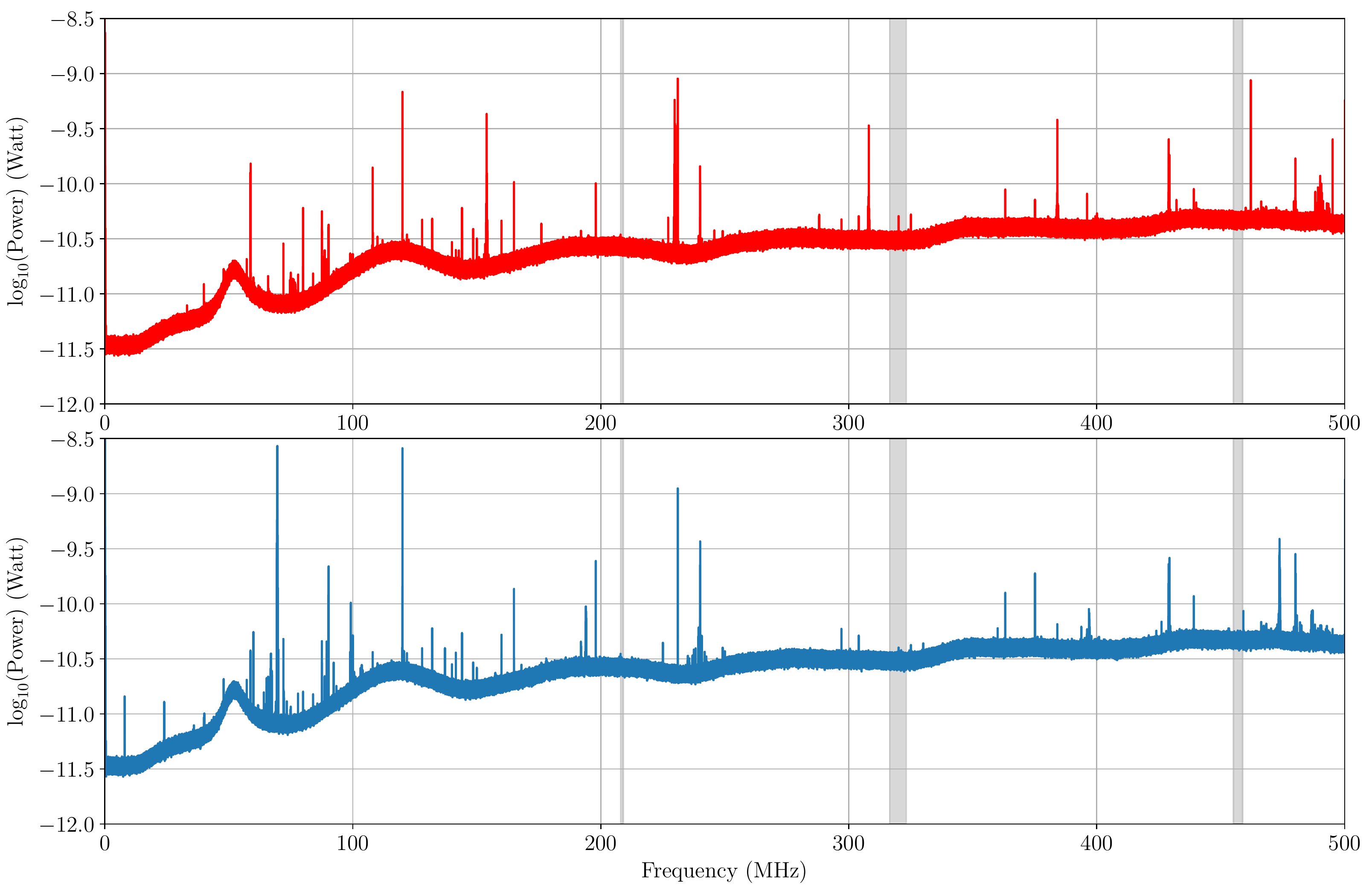}
\caption{Power spectra obtained from the first (top) and the last
  (bottom) 10-\si{\second} acquisitions of the WISPDMX measurements
  made at the initial and the final positions of the two plungers,
  respectively. The shaded areas in the panels indicate the tuning ranges
  of the resonant modes achieved during the science run.}
\label{fig:first_vs_final}
\end{figure}

Visually, the \num{22000} {\em single acquisition spectra} from the
first science run of WISPDMX have a similar power level and spectral
bandpass. The first and the last spectrum recorded over the course of
the first science run are shown in the two panels of
Fig.~\ref{fig:first_vs_final}, demonstrating the stable broadband
spectral profile. The broadband spectrum in the range below
$\approx$\SI{100}{\mega\Hz} is affected by the rapidly decreasing gain
of the amplifier chain.  In addition to this effect, there are several
prominent and time-dependent noise peaks present in the spectrum.
These noise peaks originate most likely from interference produced by
one or several devices used in the measurements (e.g., the amplifiers,
the radio switches, or the ADC clock) or from stray fields that
penetrate the cavity, even though the cavity provides strong
shielding. The stray noise may penetrate through the shielded coaxial
cables, the coaxial connectors, and the unshielded amplifier box. The
narrow band noise is subsequently amplified before recording it with
the ADC. During the first science run, we did not have the capacity to
fully investigate and eliminate the origins of these spurious
features.  Fortunately, most of them reside in the off-resonance
regions of the measured spectrum and cover multiple channels, and
hence they can be easily mitigated and excluded from the
analysis. These prominence of these interference signals should be
reduced during the planned subsequent runs of WISPDMX which will be
carried out with an improved setup and better radio shielding.

\subsection{Averaged Spectrum} \label{ave_spec}

To search for DM signal outside of the ranges covered by the four
tuned resonant modes, the {\em averaged spectrum} is generated from
the \num{22000} single acquisition spectra.  Similarly to the thermal
noise reduction achieved by the averaging process described in
Sect.~\ref{acquisition_system}, averaging of the single acquisition
spectra results in reducing the fluctuations by a factor of
$1/\sqrt{\num{22000}}$.  This is demonstrated in
Fig.~\ref{fig:lowest_power}, where the averaged spectrum and the
single acquisition spectrum are compared.  The expected reduction of
the fluctuations in the power level is clearly visible there. Apart
from this reduction of the noise fluctuation, the average spectrum is
otherwise very similar in its shape and power level to the single
acquisition spectrum.

\subsection{Signal Scan}
\label{sec:sigscan}

We scan both the \num{22000} single acquisition spectra as well as the
composite average spectrum, in order to search for a potential hidden
photon signal. The signal scan is performed in two separate steps:

\begin{itemize}

\item[(1)] {\em resonant scan}, using single acquisition spectra
  to search for signals within a narrow bandwidth centered at the
  respective resonant frequencies of the cavity modes;

\item[(2)] {\em broadband scan}, using the average spectrum to
  search for signals outside of the frequency ranges covered by the
  cavity tuning.

\end{itemize}

At each step of this analysis, we assess the noise properties,
identify potential candidates, evaluate their significance, and
calculate the specific power at which a  signal can be excluded.
For the signal scan in the resonant part of the spectrum, three
\textit{a priori} defined consecutive selection criteria are applied:

\begin{enumerate}

\item Choosing a region of interest (ROI) which is centered at the frequency $\nu_0$
  of the resonance and has a bandwidth of \SI{100}{\kilo\Hz} $\text{ROI} = [\nu_0 -
  \SI{50}{\kilo\Hz}, \nu_0 + \SI{50}{\kilo\Hz}]$. The significance level
  $\mathcal{S}$ in one channel is given by

  \begin{equation}
  \label{eqn:significance}
  \mathcal{S} = \frac{S}{\sigma}=\frac{P_\text{1 channel} - \mu_\text{ROI}}{\sigma_\text{ROI}}\,,
  \end{equation} 
  with $\mu_\text{ROI}$ and $\sigma_\text{ROI}$ are the mean power and noise
  power of the ROI respectively which are calculated using the power level of
  the channels located in the ROI\footnote{$\mu_\text{ROI} = \sum{p_i}/n$,
    $\sigma_\text{ROI} = \sqrt{\sum(p_i - \mu_\text{ROI})^2/n}$ with $n$
    channels inside the ROI and the power $p_i$ in channel $i$.}. If the power
  excess $S$ in a channel (within the ROI) has a significance $\mathcal{S} > 3$,
  the channel is recorded and becomes a \textit{level-I} signal candidate.

\item Tracking the position and amplitude of the \textit{level-I} candidate
  signal in multiple consecutive spectra: The position of the candidate signal
  must reside in the same channel while their amplitude must vary while the
  position of the resonant frequency changes in the tuning procedure.  The
  candidate which satisfies both conditions becomes a \textit{level-II}
  candidate.

\item Test the signal width condition on the \textit{level-II} candidate: if the
  number of consecutive \textit{level-II} channels is equal to the expected
  signal width of the signal in the designated frequency, the \textit{level-II}
  candidates are promoted to \textit{level-III} candidates. Their position is
  recorded for manual inspection.

\end{enumerate} 

In the narrow band searches over the on-resonance region, we are
selecting candidates both exactly on-resonance as well as slightly
off-resonance. The criterion for the \textit{level-II} candidates
introduced above takes into account that the spectral resolution
largely over-samples the width of the resonance. Following that
requirement, a true signal can be reliably discriminated against
spurious noise.  

The broadband signal search is performed using the averaged
spectrum. The conditions applied for the broadband search are limited
to satisfying the abovementioned criteria of significance (1) and
signal width (3), and the search proceeds as follows:

\begin{enumerate}[(a)]

\item Channel search (similar to the on-resonance scan): The list of
  channel candidates is retrieved after dividing the broadband
  spectrum into \num{500} ROIs with \SI{1}{\mega\Hz} width and selecting
  the channels which exceed the significance threshold of \num{5}
  $\sigma$.  
    
\item Similarly to the third step in the resonant search, the signal
  width condition is applied to select signals with consecutive
  channels with a combined width as expected for a dark matter
  signal at that particular frequency.

\end{enumerate} 

It should be noted that the signal scan on the averaged
spectrum shares many mutual features between the on- and off-resonance
scan algorithm at a single spectrum level.  However, the broadband
scan performed on the single averaged spectrum is inherently less
restrictive on identification of potential candidates because there
are no other spectra to compare (and so there is not an equivalent
of the \textit{level-II} criterion to be used for the broadband
search in the off-resonance frequency range). In this case, the
signal width criterion becomes the most important condition for
filtering out spurious candidates.

\section{Results}
\label{sec:results}
\subsection{Detectable Power and the Noise Power}

\begin{figure}
\centering 
\includegraphics[scale =0.4]{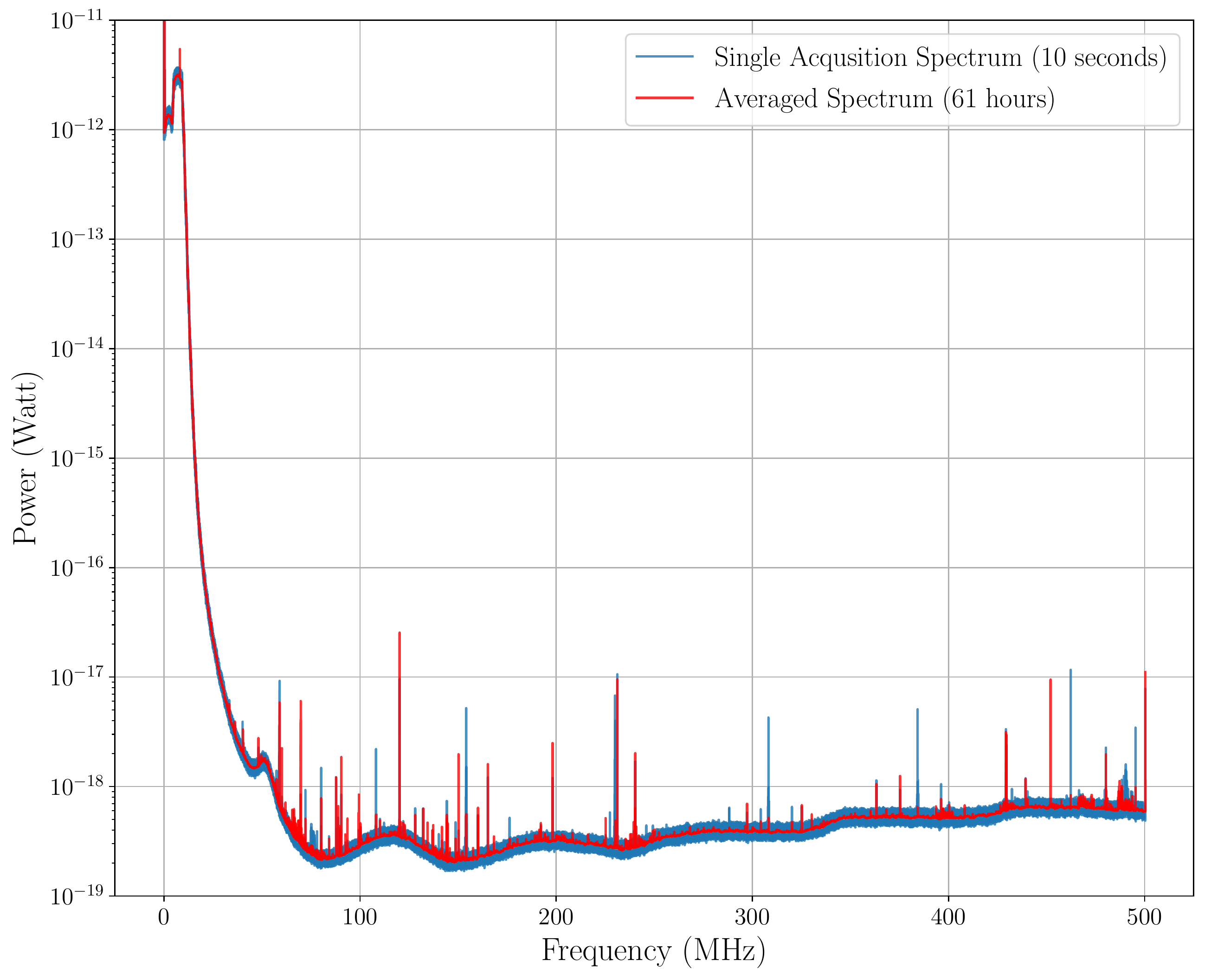}
\caption{The measured power at the antenna obtained in a single acquisition 
spectrum (gray) and after averaging the \num{22000} individual spectra (red) 
from the WISPDMX first science run.  
\label{fig:lowest_power}}
\end{figure}

The signal scan procedures described above consider a narrow band
excess power in a ROI to be signal-like if it is significant in
comparison with the fluctuations of the background spectrum and if it
matches the expected line width at the given frequency.  The measured
broadband spectrum (see Fig.~\ref{fig:lowest_power}) is dominated
by the thermal noise of the antenna, the cavity walls and (mostly)
narrow band ambient RF noise that couples into the cable connecting
the antenna to the amplifier. Additionally, the amplifier contributes
to the noise (with the given noise figure of \SI{0.518}{\decibel}). At
the digitization step, a negligible amount of noise is added to
the overall background (the \num{12}-bit digitization of the 
$\pm\SI{0.4}{\volt}$ range introduces a digitization
noise power at the level of \SI{-14}{\deci \belm}).

In the following, we combine the recorded power spectra and the
amplifier gain to estimate the measured poer at the antenna resulting from the WISPDMX
measurements and to determine the detectable excess power. As
described in Sect.~\ref{sec:amplifier}, the amplifier gain been
measured by sending a sweeping signal from the frequency generator to
the amplifier chain and recording the output power. The resulting
frequency dependent gain presented in Fig.~\ref{fig:amp_factor} is
applied to the recorded WISPDMX power spectra to calculate the
measured power spectra as illustrated in
Fig.~\ref{fig:lowest_power}. These power spectra are subsequently
used for estimating the values of $\mu_\mathrm{ROI}$ and
$\sigma_\mathrm{ROI}$ introduced in Eqn.~\ref{eqn:significance}.


The channel-by-channel fluctuation $\sigma_\text{single}$ of the
single spectrum is calculated by collecting the channel power from
\num{22000} single spectra. The $\sigma_\text{single}$ is the standard
deviation of the collection. While the noise power of the averaged
spectrum $\sigma_\text{averaged}$ is estimated from the
$\sigma_\text{single}$ by:

\begin{equation}
\sigma_\text{averaged} = \frac{\sigma_\text{single}}{\sqrt{\num{22000}}}\,,
\end{equation} 
as mentioned in Sect.~\ref{ave_spec}. The noise power at the level of the
single spectrum and averaged spectrum are $\sigma_\mathrm{single}\sim
\SI{8e-19}{\watt}$ and $\sigma_a\sim \SI{5e-22}{\watt}$.

\subsection{Results from the Signal Scan}

The resonant scan applied to the narrow band frequency intervals
selected in the single acquisition spectra around the resonant cavity
modes covers the following ranges of the particle mass
\SIrange{0.8602}{0.8646}{\micro\eV} ($\text{TM}_{010}$),
\SIrange{1.3088}{1.3358}{\micro\eV} ($\text{TE}_{111-1}$,
$\text{TE}_{111-2}$), and \SIrange{1.8820}{1.8977}{\micro\eV}
($\text{TM}_{020}$). The resulting number of \textit{level-I} signal
candidates reaching the significance level larger than 3 ($\mathcal{S}
> 3$) varies between 5 to 8, in a single ROI with $\Delta
\nu=\SI{100}{\kilo\Hz}$.  The \textit{level-I} candidates are
subsequently filtered using the position and amplitude test
performed in the consecutive spectra.  However, none of these
\textit{level-I} candidates pass the test on the expected variation of
amplitude while tuning across the resonance.  On the other hand, the
\textit{level-I} candidates are filtered using the line width
test. Compliance with the expected line width of the hidden photon
signal requires 5 consecutive channels with an excess for the
$\text{TM}_{010}$ mode, 8 consecutive channels for the twin modes
$\text{TE}_{111}$, and 11 consecutive channels for the
$\text{TM}_{020}$ mode.  In our data, there are no such
features, as none of the regions with the respective numbers of
consecutive channels has a significance level of
$\mathcal{S}>2$. This allows us to conclude that the resonant scan
searches do not detect any candidate for the hidden photon signal
above the level of $2\sigma_\text{single}$.

The broadband scan procedure applied to the average spectrum
yields a total of \num{20646} single-channel candidates with
significance level larger than 5$\sigma_\text{averaged}$. The signal
width condition is applied again to filter out the candidates and
reduces the total number of candidates down to 642 candidates,
together covering a total of 1628 channels.

\begin{table}
\caption{Number of candidates from the broadband scan satisfying the signal line width condition}
\centering
\label{tab:candi_ave}
\scalebox{0.8}{
\begin{tabular}{c|c|c||c|c|c}
\hline \hline
\textbf{Width} & \textbf{Range } & \textbf{Candidate} & \textbf{Width} 
& \textbf{Range} & \textbf{Candidate} \\
   (channel)     &  \si{\mega\Hz}  &    \textit{level-II}   & (channels) 
&  \si{\mega\Hz} & \textit{level-II}    \\ \hline \hline
  1  & 0 - 59.059         & 280  & 7   & 263.763 - 304.804  & 9  \\ \hline
  2  &  59.059 - 100.600  & 158  & 8   & 304.804 - 345.345  & 1  \\ \hline
  3  & 100.600 - 141.641  & 84   & 9   & 345.345 - 386.386  & 15 \\ \hline
  4  & 141.641 - 182.182  & 31   & 10  & 386.386 - 426.926  & 5  \\ \hline
  5  & 182.182 - 223.223  & 20   & 11  & 426.926 - 467.967  & 6  \\ \hline
  6  & 223.223 - 263.763  & 27   & 12  & 467.967 - 500      & 6  \\
\hline  
\end{tabular}}
\end{table}

The Table~\ref{tab:candi_ave} lists the number of these candidates falling
within the specific frequency ranges as determined by the expected
width of the hidden photon signal. As discussed in
Sect.~\ref{section:signaWidth}, the signal is expected to follow the
Maxwellian distribution. In a follow-up test the signal candidates are
fit with a Maxwellian signal function \cite{turner1990}. The fit is
only feasible for the candidates where the expected signal width
is resolved across at least five channels.  The frequency range used
for the fit is extended up to adjacent 10 channels to estimate the
background and to include the tail of the Maxwellian function.

\begin{figure}
  \centering 
  \includegraphics[scale =0.4]{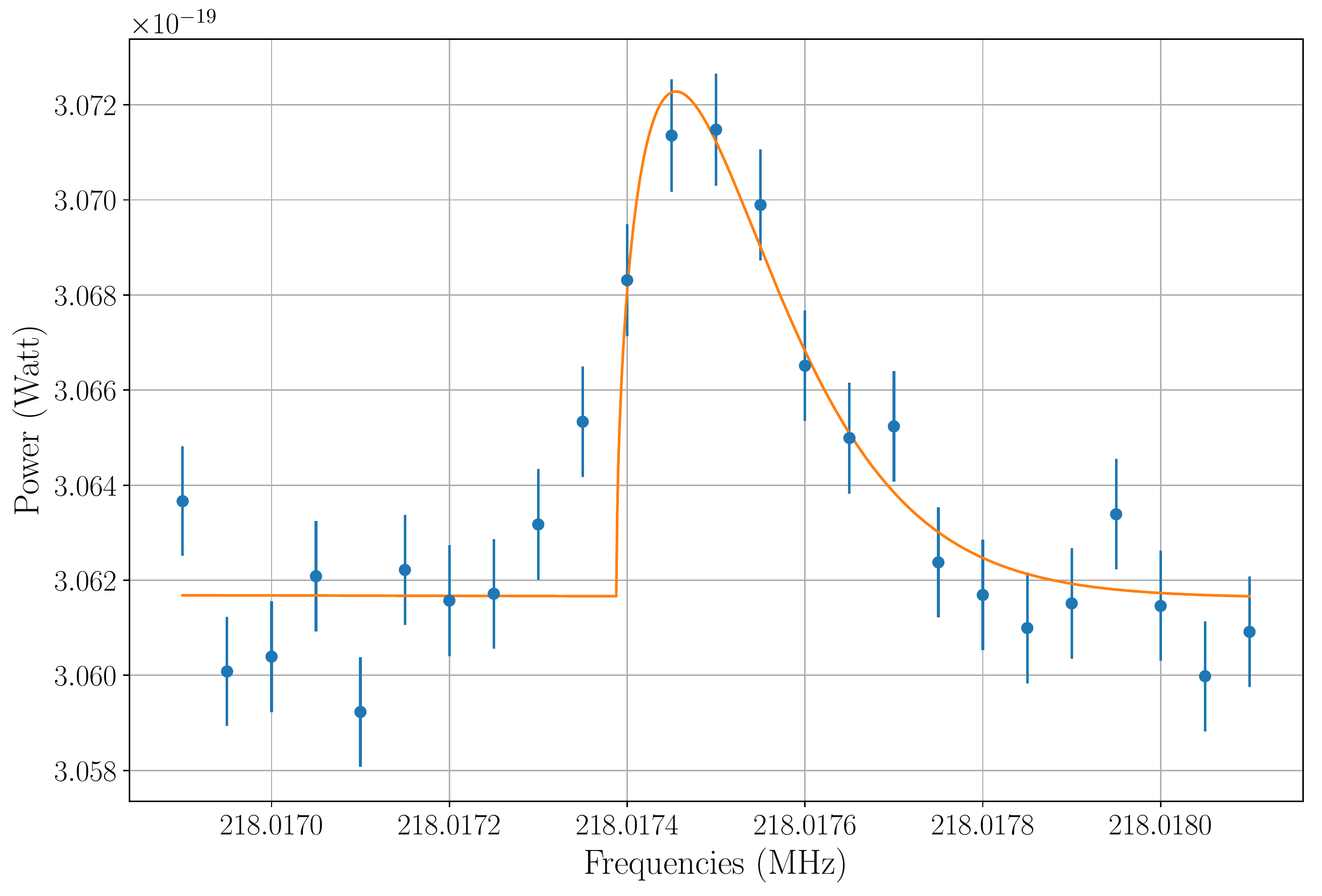}  
  \caption{A potential candidate signal with the largest $p$-value
    (8.175\%) of the fit by the Maxwellian profile.The reduced
    $\chi^2$ parameter of the fit is 1.44, calculated for 22 degrees
    of freedom. The fitted rest mass frequency of
    $218.017395(^{+9}_{-34})$\,\si{\mega\hertz} corresponds to a
    particle mass of $0.90164783(^{+4}_{-14})$\,\si{\micro\eV}. The
    width of the fitted profile translates into a dark matter halo
    velocity dispersion of
    $330^{+70}_{-60}$\,\si{\kilo\meter/\second}. If the total excess
    power of $(1.8 \pm 0.2) \times 10^{-21} \si{\watt}$ in the fitted
    profile is produced by the kinetic mixing of the hidden photons
    from the Galactic dark matter halo, the resulting coupling
    constant is $(1.1\,^{+0.6}_{-0.2}) \times 10^{-12}$.}
  \label{fig:best_candidate}
\end{figure}

The goodness-of-fit criterion is based upon the $\chi^2$ calculated
with the error given by the noise power of the averaged spectrum. Most
of the candidates can be rejected given the large $\chi^2$ values
found for the given number of degrees of freedom\footnote{The degrees
  of freedom are given by $n - m$, where $n$ is the total number of
  channels used in the fit and $m=3$ gives the number of free
  parameters in the fitting function.} and therefore small
$p$-values. The candidate with the largest $p$-value of 8.175\% is
shown in Fig.~\ref{fig:best_candidate}. For this candidate, five
channels are estimated to have a significance level larger than 5, and
the central frequency obtained from the fit is
$218.017395(^{+9}_{-34})$\,\si{\mega\Hz}.

This rest mass frequency corresponds to a particle mass of
$0.90164783(^{+4}_{-14})$ \si{\micro\eV}. The fitted width of the
Maxwellian profile corresponds to a dark matter halo velocity
dispersion of $330^{+70}_{-60}$\,\si{\kilo\meter/\second}, which
agrees well with the observational estimates
\cite{c1_Green,c1_kerr}. If the total excess power of $(1.8\pm 0.2)
\times 10^{-21}\si{\watt}$ measured in the fitted profile is produced
by the kinetic mixing of the hidden photons from the Galactic dark
matter halo, the resulting coupling constant is $(1.1\,^{+0.6}_{-0.2})
\times 10^{-12}$, which places it well within the ranges allowed for
the hidden photon cold dark matter (see,
\cite{nelson_dark_2011,arias_wispy_2012}).

However, further inspection of a broader spectral range around this
peak reveals its potential connection to a periodic feature appearing
throughout a substantial portion of the spectrum at intervals close to
\SI{333.36}{\kilo\hertz}, which are likely an instrumental effect
(e.g., an interference signal from the clock of the ADC module). None
of these quasi-periodic features has satisfied our selection criteria
for a signal candidate, with most of them being limited to two
spectral channels. We conclude nevertheless that the nature of this
candidate signal needs to be further investigated with additional
measurements that can be used to identify potential instrumental
effects and to look for the annual modulation expected for the actual
dark matter signal.

\subsection{Exclusion Limit from the First Science Run}
\label{significance}

\begin{figure}
\centering 
\includegraphics[scale=0.4]{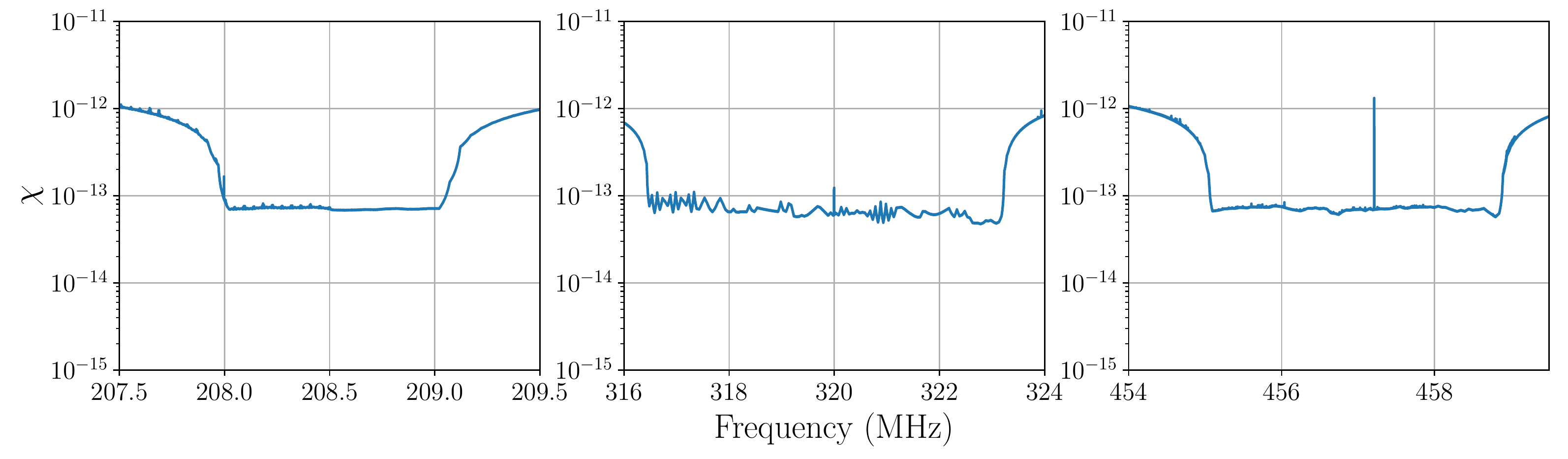}  
\caption{Exclusion limits for the kinetic mixing of hidden photons
  obtained from the resonant signal scans of WISPDMX
  measurements at a 95\% confidence level.}
\label{fig:excl_limit_3modes}
\end{figure}

\begin{figure}
\centering 
\includegraphics[scale =0.4]{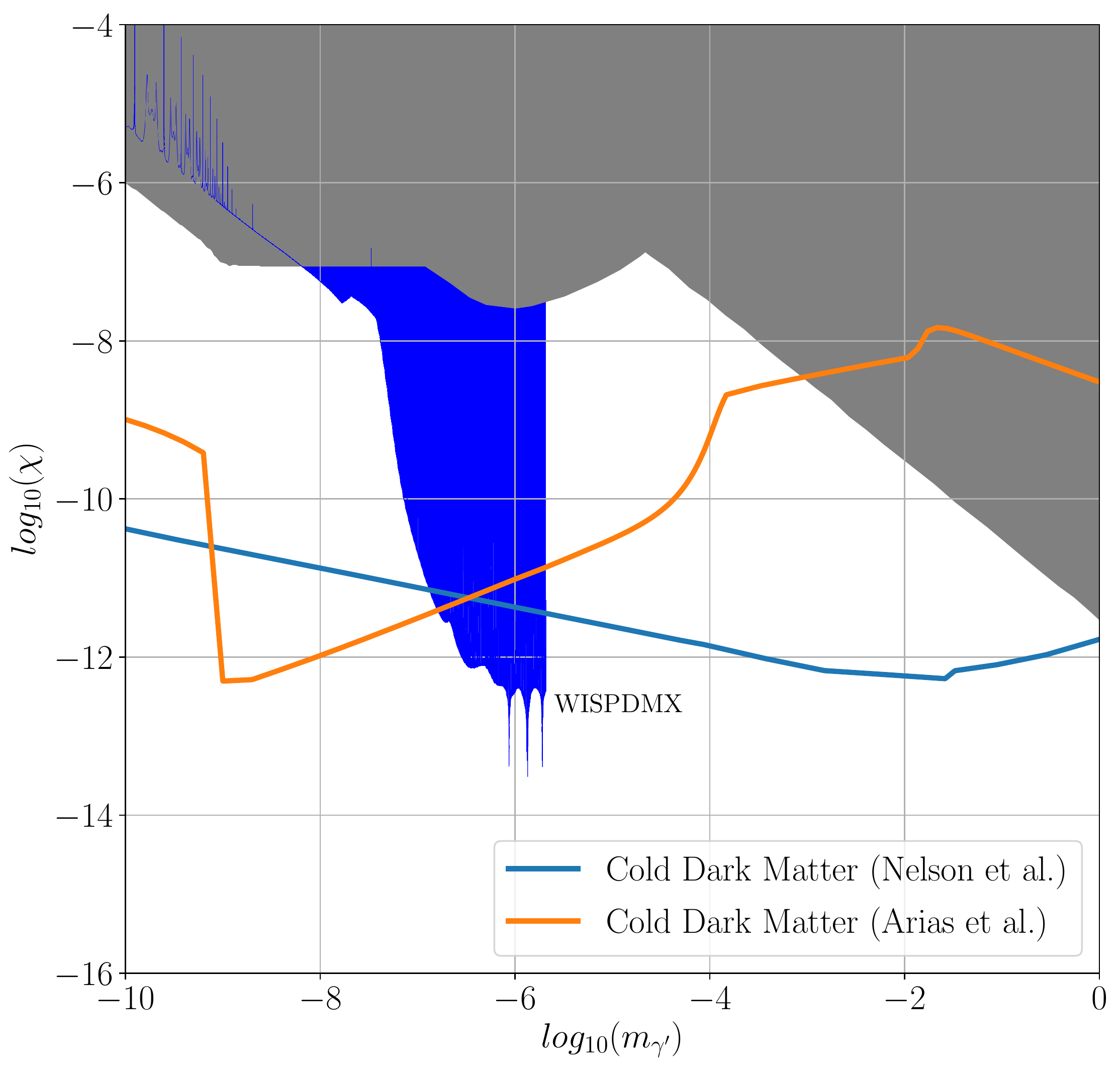} 
\caption{The combined 95\% exclusion limits for
  the hidden photon kinetic mixing obtained from the resonant and
  broadband signal scans of the WISPDMX measurements.  
   The gray shade denotes cumulative exclusion limits 
  obtained in previous experiments 
  \cite{FIRAS,c1_Georgi,c1_Mirizzi,c1_Karshenboim2}.}
\label{fig:excl_limit}
\end{figure}

Leaving the investigation of the potential signal candidate at
\SI{218.017395}{\mega\Hz} to further dedicated measurements, we proceed
here to derive limits on the coupling parameter $\chi$ over the entire
frequency range.  In case of non-detection, we can use
Eqn.~\ref{eqn:Pout} to calculate the exclusion limits for the kinetic
mixing parameter $\chi$ at a given confidence level described by the
desired signal-to-noise ratio, $S$. This yields:

\begin{equation}
\label{eqn:exclu_limit}
\chi = 6.41 \times 10^{-5} \sqrt{S}\, \left(\frac{\sigma_\mathrm{noise}}{[W]}\right)^{1/2} 
\left(\kappa(\nu)\, g(\nu)\, \frac{V}{[\si{\liter}]}\, \frac{m_{\gamma'}}{[\si{\eV}]}\, 
\frac{G_\rho\, \rho_\mathrm{DM}}{[\si{\giga\eV\,\cm^{-3}}]} \right)^{-1/2}\,.
\end{equation} 
The measured antenna coupling, $\kappa(\nu)$, is used for the calculations. 
The measurement gain, $g(\nu)$ is calculated separately for each individual
spectrum analyzed. The calculations of $\chi$ are done assuming an isotropic
directional distribution of the incident hidden photons and using
$\rho_\mathrm{DM} = \SI{0.3}{\giga\eV\,\cm^{-3}}$ and $V=\SI{447}{\liter}$.
Figure~\ref{fig:excl_limit_3modes} presents the resulting 95\%
exclusion limits obtained in the resonant regime within the tuning
ranges of the individual cavity modes. In Fig.~\ref{fig:excl_limit},
these exclusion limits are amended with the broadband exclusion
limits obtained from the averaged spectrum, and the combined limits
are compared with the parameter space excluded by other experiments
(see \cite{arias_wispy_2012,essig_dark_2013} for details). The figure
also shows theoretical predictions for the kinetic mixing allowed for
hidden photons constituting dark matter
\cite{nelson_dark_2011,arias_wispy_2012}.

Figure~\ref{fig:excl_limit} shows that the WISPDMX measurements have
improved the existing exclusion limits in a broad range of hidden
photon mass, from \SI{0.4}{\nano\eV} to \SI{2.07}{\micro\eV}.  Above
$\approx 0.1\si{\micro\eV}$, the WISPDMX has excluded part of the parameter
space for the hidden photon dark matter predicted in the misalignment
Skotinogenesis scenarios \cite{nelson_dark_2011,arias_wispy_2012}.


\section{Discussion}
\label{sec:discussion}

The measurements described in this paper comprise the first fully
automated science run made with the WISPDMX apparatus, achieving a
substantial improvement in comparison to the earlier trial runs made
in the untuned \cite{baum2013,horns2014} and partially tuned
\cite{nguyen2015} modes of operation.
The present measurements
extend the existing haloscope exclusion limits for hidden photon dark
matter (calculated in \cite{arias_wispy_2012} using the results from
several axion dark matter searches
\cite{bradley2003,asztalos2004,asztalos2010}) to masses below
\SI{2.07}{\micro\eV} and pioneer a combination of resonant and
broadband signal search, extending the search range down to
\SI{0.4}{\nano\eV}. 

For the resonant modes, the sensitivity and exclusion limits can
potentially be improved by applying the multiple bin analysis used in 
\cite{asztalos2001}. The quality of the broadband spectrum can be
improved if the measurements are repeated in a better shielded
environment which should help reducing the interference from external
radio frequency signals, especially within the FM radio band at
\SIrange{80}{110}{\mega\hertz}. 

A dedicated multiple-run arrangement in which measurements at each
given cavity tuning performed at three different local sidereal times
would further refine the results by improving the experiment
sensitivity to local dark matter flows with non-isotropic directional
distribution of the particle velocity \cite{arias_wispy_2012}. With
the cavity positioned at a 45$^\circ$ angle with respect to the
rotational axis of the Earth, one of these three measurements should
fall with a factor of 0.75 of the sensitivity expected for the
isotropic directional distribution.

The nature of the signal candidate found at \SI{218.017395}{\mega\Hz}
(\SI{0.90164783}{\micro\eV}) needs to be better understood. The halo
velocity dispersion of \SI{330}{kilo\meter\,\second^{-1}} inferred
from this signal agrees well with the observational estimates
\cite{c1_Green,c1_kerr}. The total excess power of $(1.8\pm0.2) \times
10^{-21}\si{\watt}$ measured in the fitted profile can be reconciled
with the estimated energy density of the Galactic dark matter halo
composed of hidden photons kinetically mixed with normal photons at
$\chi = (1.1\,^{+0.6}_{-0.2}) \times 10^{-12}$.  Thus the shape, the
width, and the strength of this signal candidate can reconcile it with
the signal from the hidden photon dark matter
\cite{nelson_dark_2011,arias_wispy_2012}.  However, the physical
relevance of this signal still needs to be investigated further,
particularly with respect to its potential relation to a periodic
interference signal identified throughout a broader frequency range in
the measured spectrum.

Such an investigation will be realized by repeating the WISPDMX runs
with a better shielding, an improved apparatus, and at different
times of the year. The first two measures
would provide a better account of potential instrumental effects,
while the latter one should constrain the physical relevance of the
signal by looking for its expected annual modulation.  With the
parameters of the tentative signal, the expected frequency shift of the signal
due to the annual
modulation should be $\approx 16.0\si{\hertz}$, which can be detected
already at the present frequency resolution of \SI{50}{\hertz} and the
signal strength ($\mathrm{SNR}\approx 10$).

The overall results obtained with the WISPDMX so far demonstrate the
potential of employing a haloscope setup to unveil the nature of
hidden photon dark matter by combining signal detection at multiple
resonant mode and non-resonant detection over a broad range of
frequency. Such a combination has become possible due to
implementation of a high resolution, broadband recording setup which
provides a frequency resolution of \SI{50}{\Hz} over a
\SI{500}{\mega\Hz} bandwidth.

The high-efficiency acquisition system of WISPDMX provides a
methodological foundation for future broadband experiments aimed at
searching for either the hidden photon or the axion dark matter.  
The WISPDMX measurements also pave the way to
generic heterodyne experiments at the particle masses below
\SI{1}{\micro\eV}, where both resonant
\cite{chaudhuri2015,chaudhuri2018} and inherently broadband setups can
be employed.

\acknowledgments

This research was funded by the Deutsche Forschungsgemeinschaft (DFG,
German Research Foundation) within the framework of Germany's
Excellence Strategy - EXC 2121 ``Quantum Universe'' - 390833306. The
authors acknowledge the support from the Collaborative Research Center
(Sonderforschungsbereich) SFB 676 ``Particles, Strings, and the Early
Universe'' funded by the German Research Society (Deutsche
Forschungsgemeinschaft, DFG). The experiment was partially supported
through a PIER Ideenfonds grant. We thank the DESY staff for making
the HERA cavity available for the WISPDMX measurements and providing
initial technical assistance with the setup.

\bibliographystyle{JHEP}
\bibliography{wispdmx}
\end{document}